\documentclass[a4paper,11pt]{article}
\pdfoutput=1

\usepackage[a4paper,left=2.73cm,right=2.7cm,top=3cm,bottom=3.5cm]{geometry}

\usepackage{amsmath,amssymb,graphicx,wrapfig,caption,subcaption,enumerate,color,tikz}
\usetikzlibrary{arrows}
\usepackage{colortbl}
\usepackage{booktabs}

\usepackage[mathscr]{euscript}

\usepackage[colorlinks=true,linktocpage=true,linkcolor=blue,citecolor=blue]{hyperref}

\definecolor{gray}{rgb}{.9,.9,.9}
\tikzstyle{block} = [rectangle, text width=5em, text  centered, rounded corners]
\tikzstyle{line} = [draw, very thick, -latex']
\tikzstyle{RGflow}= [->, shorten <=1pt, thick, dashed, color=black!70]

\addtocontents{toc}{\protect\setcounter{tocdepth}{2}}
\numberwithin{equation}{section}

\newcommand{\mt}[1]{\textrm{\tiny #1}}
\newcommand{\sac}{\, , \qquad}
\newcommand{\eqq}[1]{(\ref{#1})}
\newcommand{\fig}[1]{Fig.~\ref{#1}}

\renewcommand{\xi}{\overline N_\mt{q}}

\newcommand{\be}{\begin{equation}}
\newcommand{\ee}{\end{equation}}
\newcommand{\bsal}{\begin{align}}
\newcommand{\bal}{\begin{aligned}}
\newcommand{\eal}{\end{aligned}}
\newcommand{\bse}{\begin{subequations}}
\newcommand{\ese}{\end{subequations}}

\begin{document}

\begin{titlepage}

\thispagestyle{empty}

\begin{flushright}
\hfill{ICCUB-18-008}
\end{flushright}

\vspace{40pt}  
	 
\begin{center}

{\LARGE \textbf{Heating up Exotic RG Flows with Holography}}
	\vspace{30pt}
		
{\large \bf Yago Bea$^{1}$  and David Mateos$^{1,\,2}$}

\vspace{25pt}

{\normalsize  $^{1}$ Departament de F\'\i sica Qu\'antica i Astrof\'\i sica and Institut de Ci\`encies del Cosmos (ICC),\\  Universitat de Barcelona, Mart\'\i\  i Franqu\`es 1, ES-08028, Barcelona, Spain.}\\
\vspace{15pt}
{ $^{2}$Instituci\'o Catalana de Recerca i Estudis Avan\c cats (ICREA), \\ Passeig Llu\'\i s Companys 23, ES-08010, Barcelona, Spain.}\\

\vspace{40pt}
				
\abstract{
We use holography to study finite-temperature deformations of RG flows that have exotic properties from an RG viewpoint. The holographic model consists of five-dimensional gravity coupled to a scalar field with a potential. Each negative extrema of the potential defines a dual conformal field theory.  We find all the black brane solutions on the gravity side and use them to construct the thermal phase diagrams of the dual theories. We find an intricate phase structure that reflects and extends the exotic properties at zero temperature.
}

\end{center}

\end{titlepage}

\tableofcontents

\hrulefill
\vspace{10pt}

%%%%%%%%%%%%%%%
%%%%%%%%%%%%%%%%%%%%%%%%%%%%%%%%
\section{Introduction}
\label{intro}
%%%%%%%%%%%%%%%%%%%%%%%%%%%%%%%%
%%%%%%%%%%%%%%%
%%%%%%%%%%%%%%%
%%%%%%%%%%%%%%%%%%%%%%%%%%%%%%%%
Holography provides a valuable tool to study non-perturbative properties of strongly coupled Quantum Field Theories (QFT). In particular, it ``geometrizes'' the Renormalization Group (RG) in the sense that it maps the properties under changes of scale of a QFT to the properties of some dual geometry in one dimension higher than the QFT. 

In this paper we will focus on thermal properties of QFTs that arise as deformations of some Conformal Field Theory (CFT) and whose RG flows exhibit exotic properties from the QFT viewpoint \cite{Kiritsis:2016kog}.\footnote{See \cite{Gursoy:2016ggq} for an earlier example.} Our model  consists of five-dimensional gravity coupled to a scalar field whose potential possesses several negative extrema (but possibly also some positive ones). Each negative extremum gives rise to an AdS solution which is dual to a CFT. 

\begin{figure}[t!!!]
\centering
\includegraphics[width=.495\textwidth]{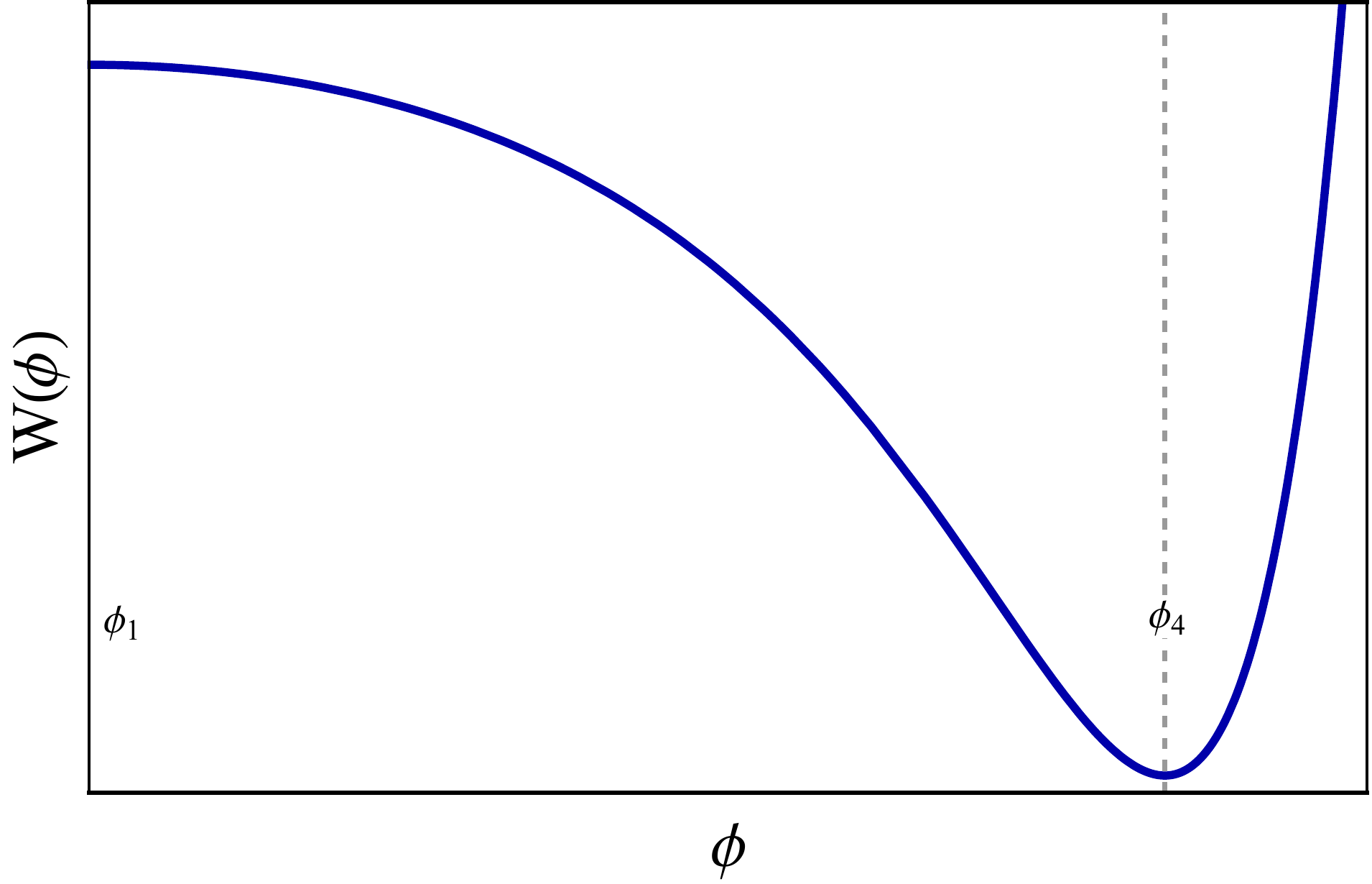}
\includegraphics[width=.495\textwidth]{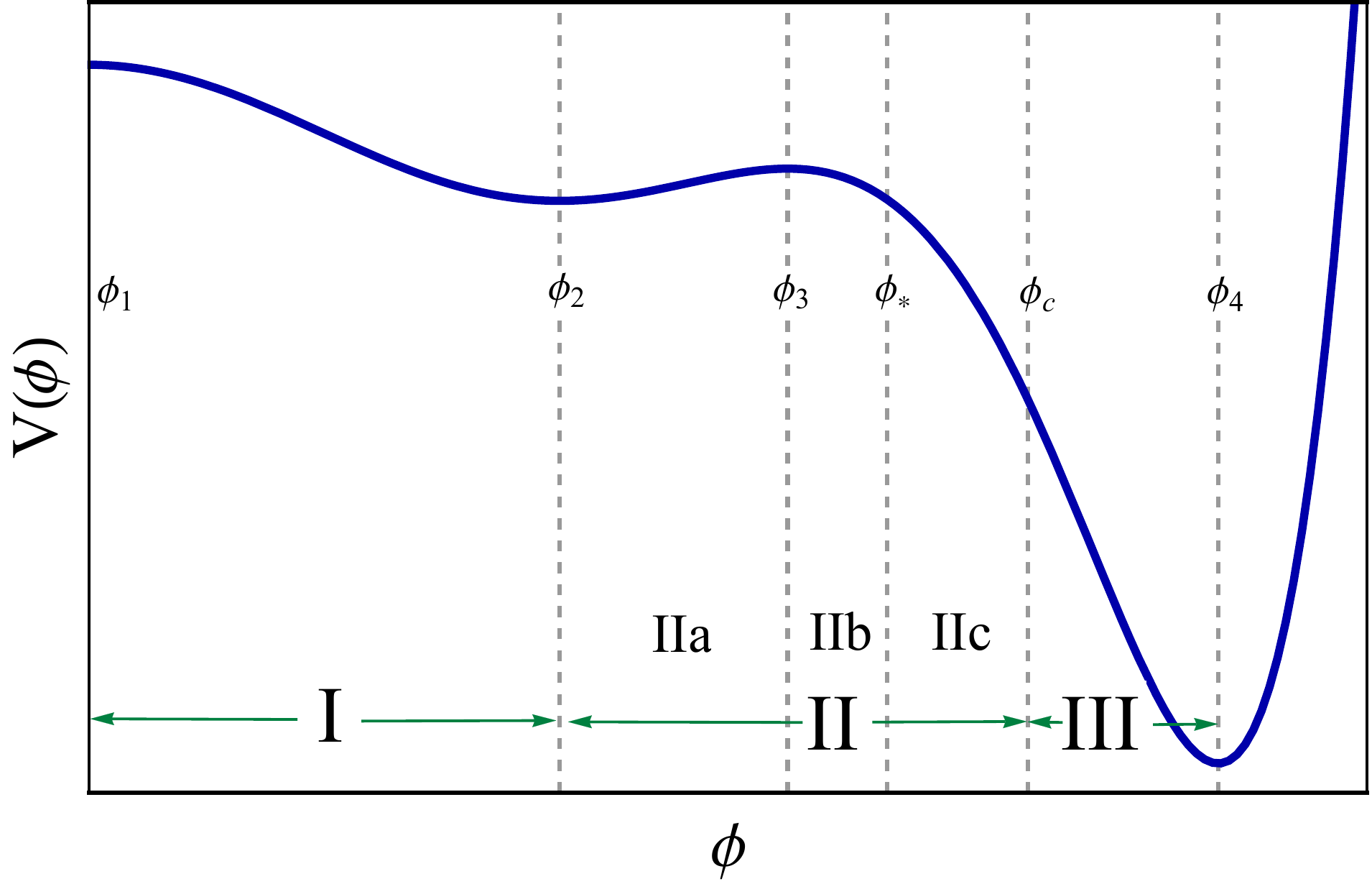}
\caption{\label{fig:Potential_qualitative} \small
Qualitative forms of the superpotential (left) and potential (right) of our model. The definition of the points $\phi_*$, $\phi_c$ and the different regions labelled I, II and III will be explained around Eqs.~\eqq{def1} and \eqq{def2}.}
\end{figure}

The qualitative form of our potential is depicted on the right-hand side of  \fig{fig:Potential_qualitative} (the case with positive extrema will be discussed in Section \ref{positive_potential_section}). This potential can be derived from a superpotential whose qualitative form is shown in \fig{fig:Potential_qualitative}(left). Given a purely bosonic theory with a potential, the superpotential is not uniquely defined. However,  we want to mimic  a situation in which our model is the bosonic truncation of a truly supersymmetric theory, in which case there would be a preferred superpotential. Therefore, as part of the definition of our toy model, we will  imagine that the ``true'' superpotential is the one in \fig{fig:Potential_qualitative}(left). Under this assumption, the extrema of the potential that are also extrema of the superpotential, labelled $\phi_1$ and $\phi_4$ in \fig{fig:Potential_qualitative},   would be dual to supersymmetric CFTs, whereas the extrema labelled $\phi_2$ and $\phi_3$ in \fig{fig:Potential_qualitative}(right)  would be dual to  non-supersymmetric ones.  In what follows we will continue to use this ``supersymmetric" versus ``non-supersymmetric" terminology with the understanding that it is meaningful only in reference to our choice of superpotential.  

Our goal will be to construct all the black brane solutions of the gravity model and to map each solution to a thermal state in one of the CFTs. As we will see, the fact that this map is non-trivial will result in interesting features of the phase diagrams of the dual CFTs.  It is not surprising that some of these features resemble those found in the case of CFTs defined on a curved space \cite{Ghosh:2017big}, since both the temperature and the boundary curvature act as infrared (IR) cut-offs in the CFT.

We will see that the flows at non-zero temperature reflect but also extend some of the exotic properties of the zero-temperature flows. By smoothly deforming the potential on the gravity side we will show that, in some cases, the exotic thermal phase structure of the dual field theories can be continuously connected with more familiar non-exotic cases.  We will also see that some of the exotic structures persist in cases in which the potential on the gravity side develops de Sitter-like maxima with positive energy density. 
\newline
\newline
{\bf Note added:} While this paper was being typewritten we became aware of Ref.~\cite{aware}, which has significant overlap with our results.

%%%%%%%%%%%%%%%%%%%%%%%%%%%%%%%%
%%%%%%%%%%%%%%%
%%%%%%%%%%%%%%%
%%%%%%%%%%%%%%%%%%%%%%%%%%%%%%%%

\section{The model}
\begin{figure}
\centering
\includegraphics[width=.5\textwidth]{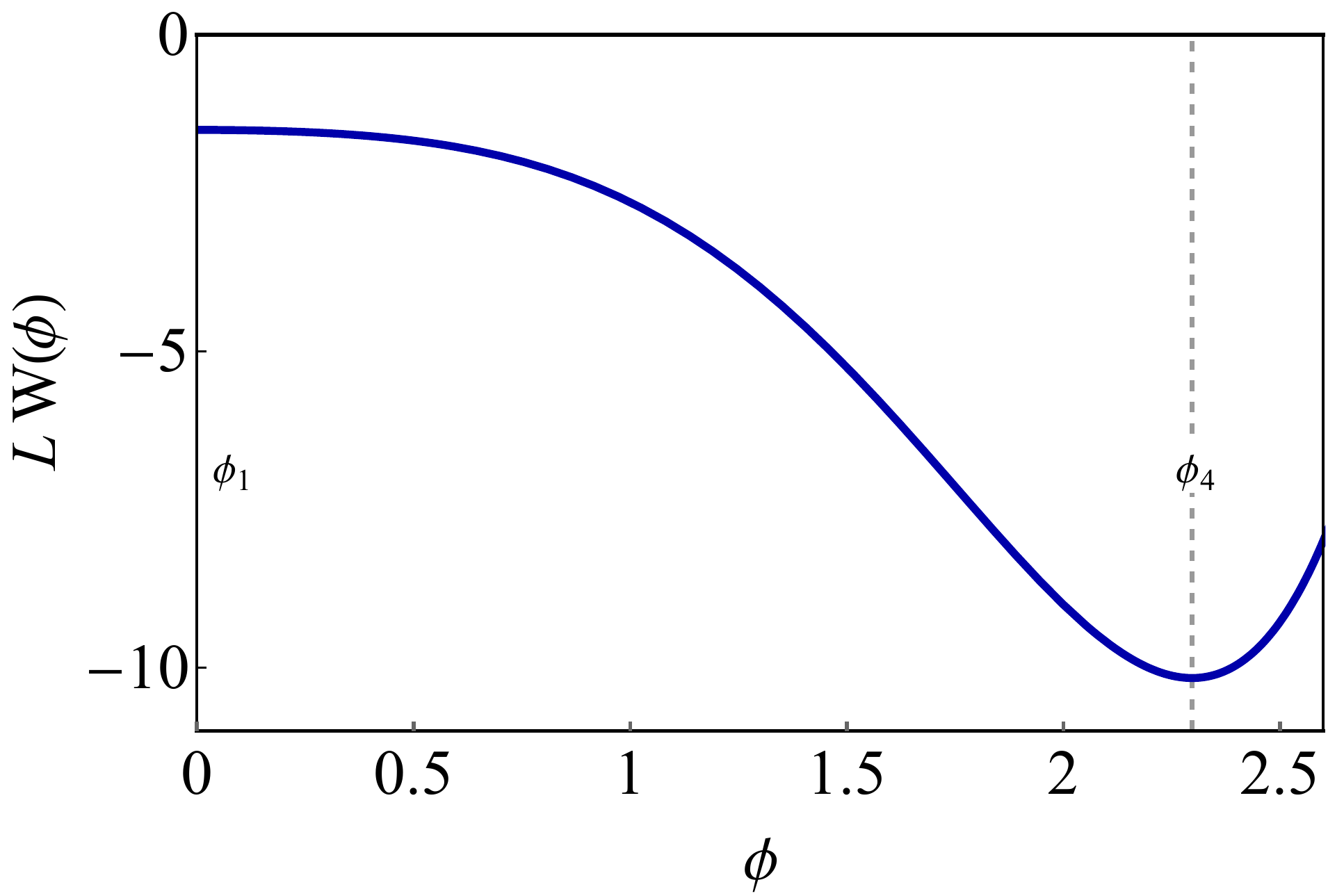}
\caption{\label{fig:superpotential} \small
Superpotential of our model.}
\end{figure}

\begin{figure}[t!!!]
\centering
\includegraphics[width=.495\textwidth]{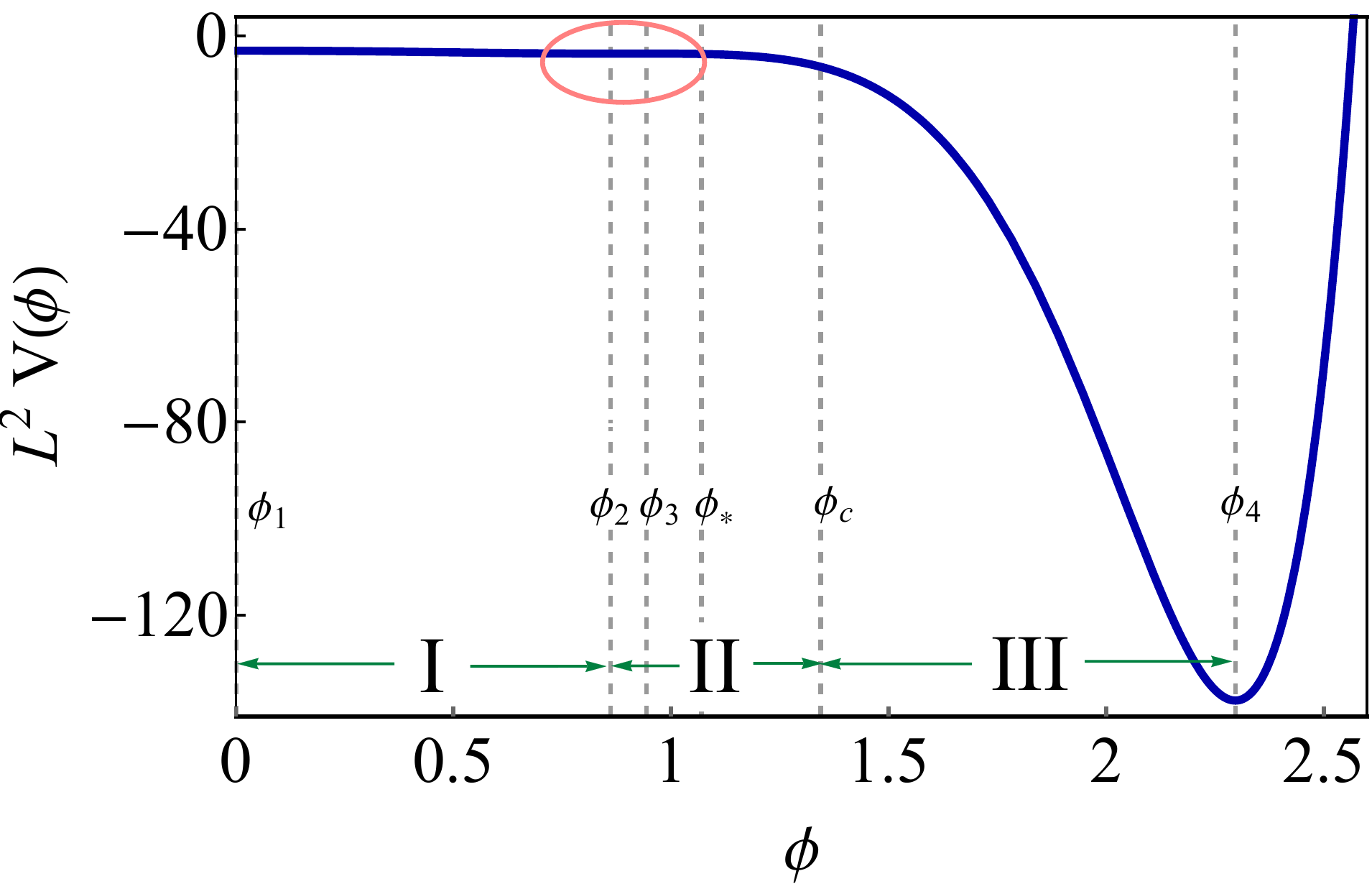}
\includegraphics[width=.495\textwidth]{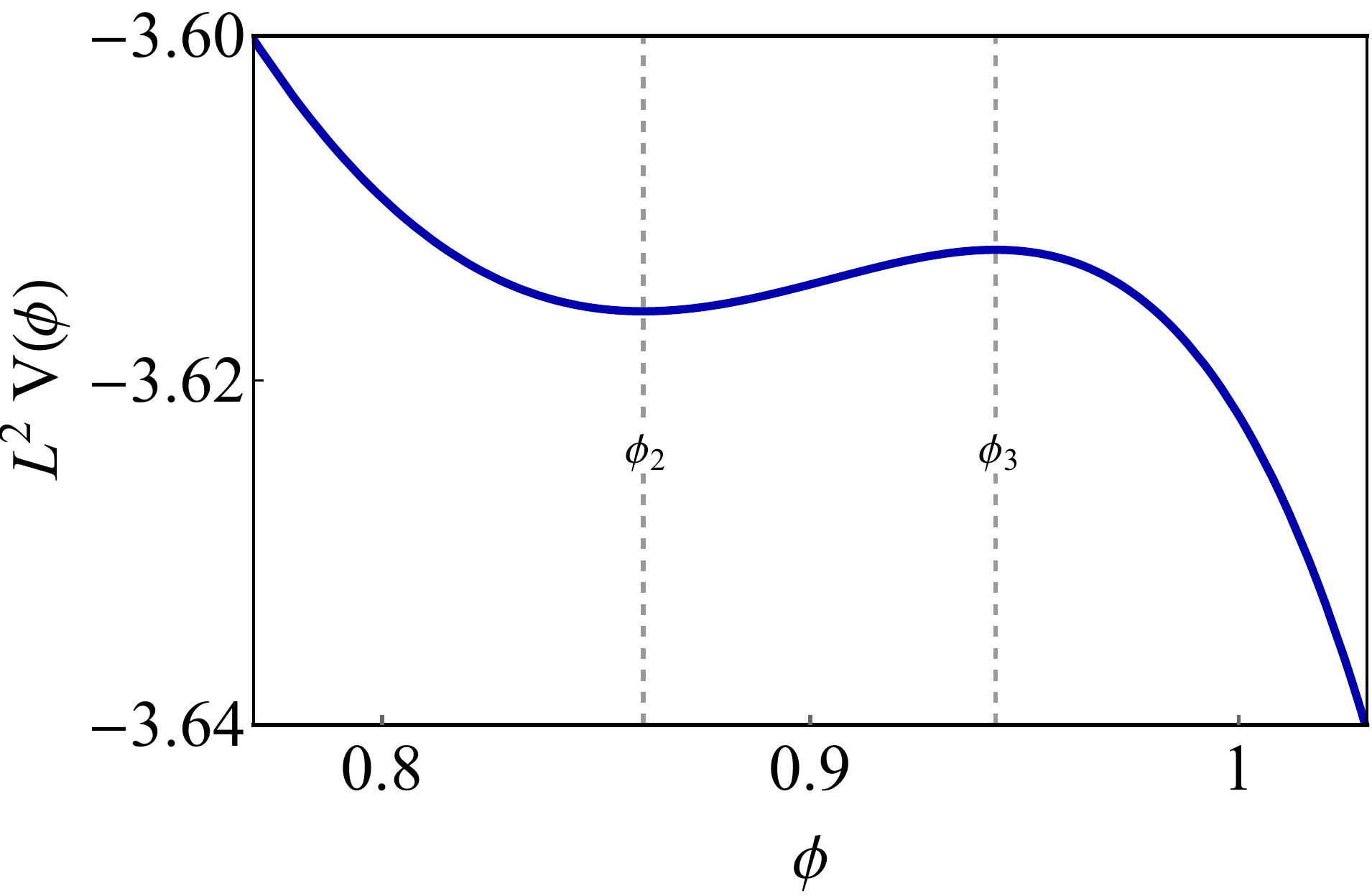}
\caption{\label{fig:potential} \small
Potential of our model. The definition of the points $\phi_*$, $\phi_c$ and the different regions labelled I, II and III will be explained around Eqs.~\eqq{def1} and \eqq{def2}. On the right we zoom into the region marked with a red circle on the left. 
}
\end{figure}

We study the Einstein-scalar model with action
\begin{equation}
\label{eq:action}
S=\frac{2}{\kappa_5^2} \int d^5 x \sqrt{-g} \left[ \frac{1}{4} {\cal R}  - \frac{1}{2} \left( \nabla \phi \right) ^2 - V(\phi) \right ] ,
\end{equation}
with a potential $V(\phi)$ derived from a superpotential $W(\phi)$ through 
\begin{equation}
V(\phi)=-\frac{4}{3}W(\phi)^2+\frac{1}{2}W'(\phi)^2~~.
\label{potential_from_superpotential}
\end{equation}
By taking derivatives with respect to $\phi$ on both sides of  (\ref{potential_from_superpotential}) we see that an extremum of $W(\phi)$ will automatically be an extremum of $V(\phi)$, but the coverse is not true in general. We consider a particular model where the potential has more extrema than the superpotential. Our superpotential is:
\begin{equation}
L W(\phi)=-\frac{3}{2}-\frac{\phi^2}{2} - \frac{\phi^4}{4 \phi_M^2} +\frac{\phi^6}{\phi_Q}~~,
\label{superpotential}
\end{equation}
where $L$ is a length scale, and we choose $\phi_Q = 10 $ and $\phi_M  \simeq 0.5797  $. With these values for the parameters the superpotential has a maximum at $\phi_1=0$ and a minimum at 
$\phi_4\simeq 2.297$, as shown in Fig.~\ref{fig:superpotential}. 
These points also correspond to supersymmetric extrema of the potential, which is displayed in Fig.~\ref{fig:potential}. 
In addition, the potential possesses  a non-supersymmetric minimum at $\phi_2 \simeq 0.861$ and a non-supersymmetric maximum at $\phi_3 \simeq 0.943$, see Fig.~\ref{fig:potential}(right). At the extrema $\phi_1,\phi_2,\phi_3,\phi_4$ of the potential  the theory admits AdS solutions with radii given by 
\be
L_1=L \sac  L_2\simeq0.9109 L \sac L_3\simeq 0.9113 L 
\sac L_4\simeq0.1476 L \,.
\ee
We will denote  the  dual conformal field theories  by $\text{CFT}_1$, $\text{CFT}_2$, $\text{CFT}_3$, $\text{CFT}_4$. 
For convenience, we have chosen the specific values of the superpotential parameters $\phi_Q , \phi_M$ so  that the mass of the scalar field is $m_1^2 L_1^2 =m_3^2 L_3^2=-3$ both at $\phi_1$ and $\phi_3$, which implies that the dual scalar operator has mass dimension three. For concreteness, in this paper we will restrict our attention to black brane solutions for which the value of the scalar field at the horizon, $\phi_H$, lies between $\phi_1$ and $\phi_4$. Since the potential is invariant under $\phi \rightarrow -\phi$, there is no loss of generality in considering only positive values of $\phi$. Moreover, a preliminary exploration  indicates that none of the physics that we will discuss is affected by the form of the potential beyond $\phi_4$, in particular by the presence of an extra 
non-supersymmetric maximum at $\phi_5 \simeq 4.130$.

As mentioned above, the form of our potential near $\phi_1$ and $\phi_3$ 
describes explicit deformations of the $\text{CFT}_1$ and the 
$\text{CFT}_3$ by a source $\Lambda$ for a dimension-three scalar operator. We will also consider deformations of any of the CFTs by expectation values 
$\langle \mathcal{O} \rangle$ of the corresponding operator. Although these are not expectation values  in the vacuum but in thermal states, in an abuse of language we will still refer to them as VEVs.  Our goal will be to find all the black brane solutions in the model with $\phi_1 \leq \phi_H \leq \phi_4$ and to use them to construct the phase diagrams of the dual CFTs.

%%%%%%%%%%%%%%%%%%%%%%%%%%%%%%%%
%%%%%%%%%%%%%%%
%%%%%%%%%%%%%%%
%%%%%%%%%%%%%%%%%%%%%%%%%%%%%%%%

\section{Study of the flows}
\label{solutions}

We use the following ansatz for black brane solutions:
\begin{equation}
ds^2_5=e^{2A(r)}\Big[ -h(r) dt^2 + d\bold{x}^2 \Big] +\frac{dr^2}{h(r)}~~.
\label{ansatz}
\end{equation}
The scalar field is also a function of $r$. The reparametrization  freedom in the $r$ direction means that the solution is completely specified by two functions instead of three, for example by  giving $A(\phi)$ and $h(\phi)$. 
  
The solutions we seek  are regular at and outside the horizon and asymptote to AdS$_5$ at large $r$. By studying these solutions we can reconstruct the thermodynamics of the dual field theories.

\subsection{$\text{CFT}_1$}
\label{subsectionCFT1}

In this section we will construct thermal states of the $\text{CFT}_1$ deformed by a source $\Lambda$ for the dimension-three scalar operator $\mathcal{O}$ dual to $\phi$. These states  are dual to gravity solutions for which $\phi$ approaches $\phi_1$ asymptotically. 

Let us start by considering the vaccum of the theory. Since the potential is derived from a superpotential, we can solve the BPS equations to find a solution connecting $\phi_4$ and $\phi_1$, acording to the structure of the superpotential, see \fig{fig:superpotential}. This solution interpolates smoothly between two AdS$_5$ geometries of different radii.  It is  known as a ``skipping'' solution because it skips the non-supersymmetric extrema at  
$\phi_3$ and $\phi_2$ \cite{Kiritsis:2016kog}. The state in the $\text{CFT}_1$  dual to this flow has zero temperature because the gravity solution has no horizon, and it has zero energy because we use a supersymmetry-preserving holographic renormalization scheme.

\begin{figure}[tbhp]
\centering
\includegraphics[width=.495\textwidth]{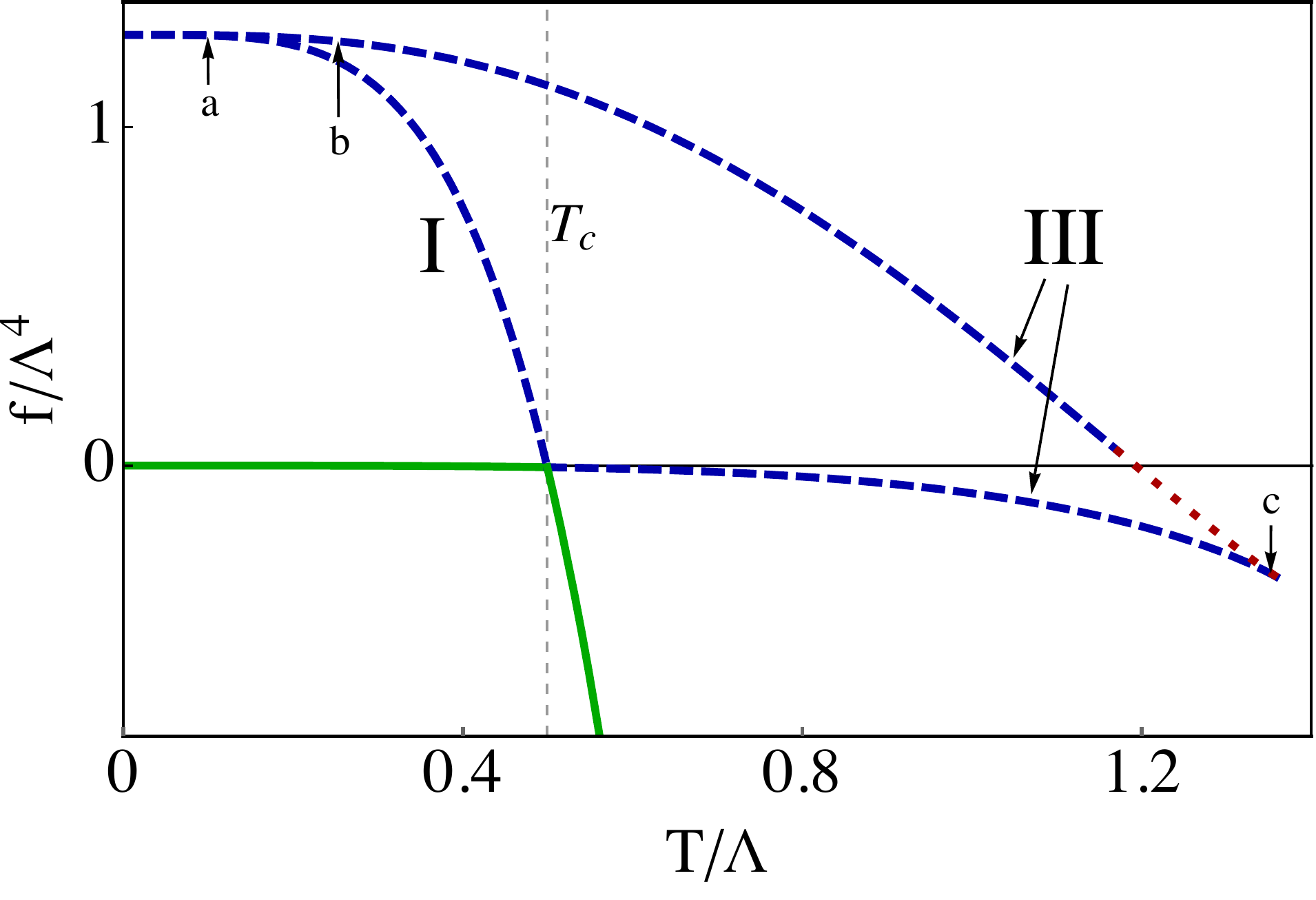}
\includegraphics[width=.495\textwidth]{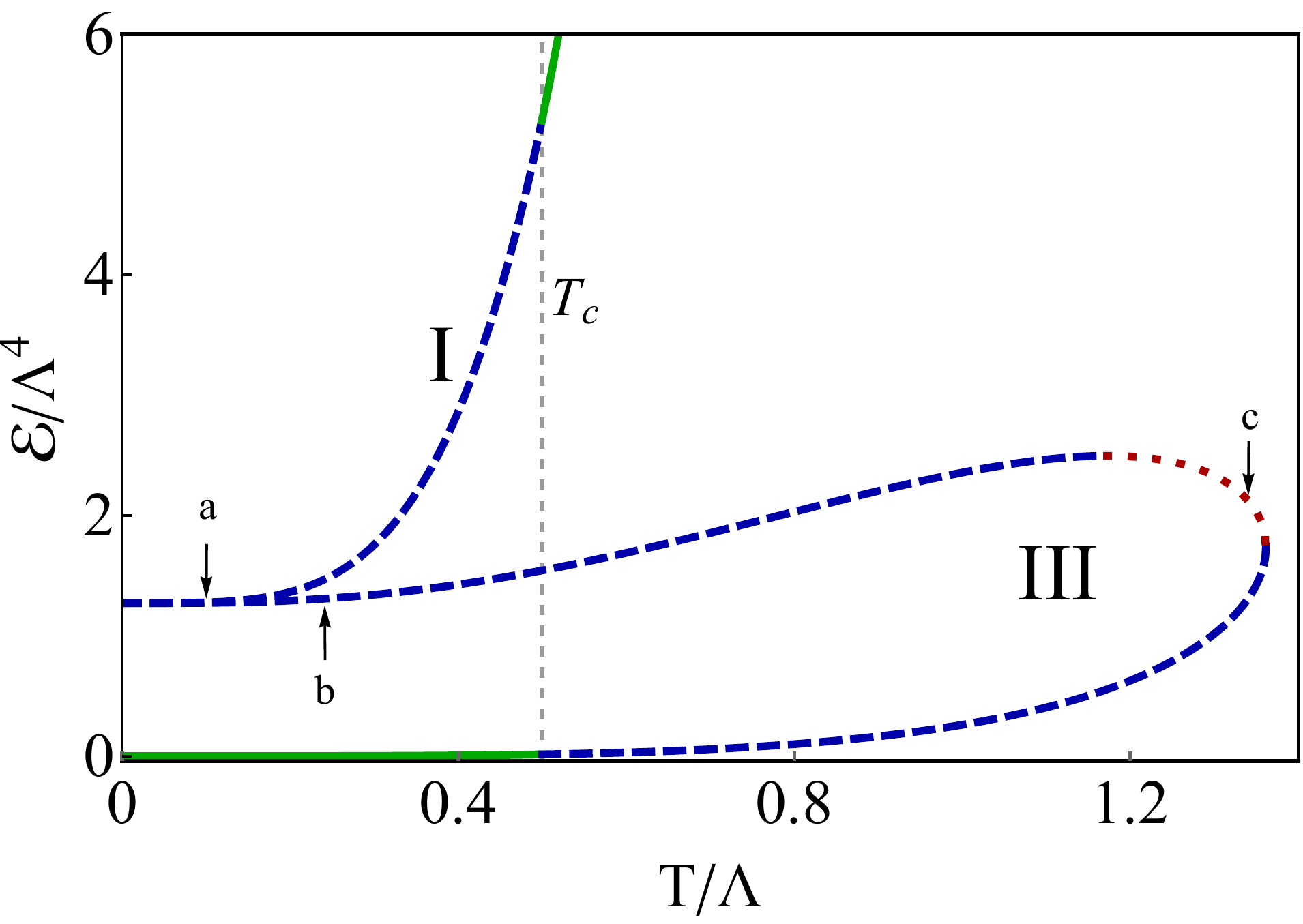}
\caption{\label{phi0_free_energy_energy}  \small
Phase diagram of the $\text{CFT}_1$. Left: Free energy density versus temperature. Right: Energy density versus temperature. The  dashed vertical line corresponds to the phase transition temperature  $T_c/\Lambda\simeq 0.499$. Locally unstable regions are shown in dotted red, locally stable but globally unstable regions are shown in dashed blue, and globally thermodynamically preferred regions are shown in solid green. The labels ``a", ``b" and ``c"  indicate the thermal states whose dual gravity solutions are displayed in \fig{Flows_examples_FT1} }
\end{figure}

The $\text{CFT}_1$ theory has a second Lorentz-invariant state to which we will refer as a second vacuum.  It corresponds to another horizonless gravity solution connecting $\phi_1$ and $\phi_2$, which again smoothly interpolates between the corresponding  AdS$_5$ geometries. However, in this case the minimum of the potential at  $\phi_2$ is not a minimum of the superpotential, so this solution is not supersymmetric.  The dual field theory state has zero temperature but strictly positive energy density. In \fig{phi0_free_energy_energy}  these two vacua correspond to the two points where the curves touch the vertical axis. 
Note that the energy difference between them is independent of the renormalization scheme. Moreover, this difference is consistent with our choice of terminology regarding ``supersymmetric" and ``non-supersymmetric" solutions in the sense that the energy of the non-supersymmetric vacuum is higher than that of the supersymmetric one. In this language, the non-supersymmetric vacuum provides a simple example of metastable dynamical supersymmetry breaking \cite{Intriligator:2006dd}. 

\begin{figure}[t!!!]
\centering
\includegraphics[width=.49\textwidth]{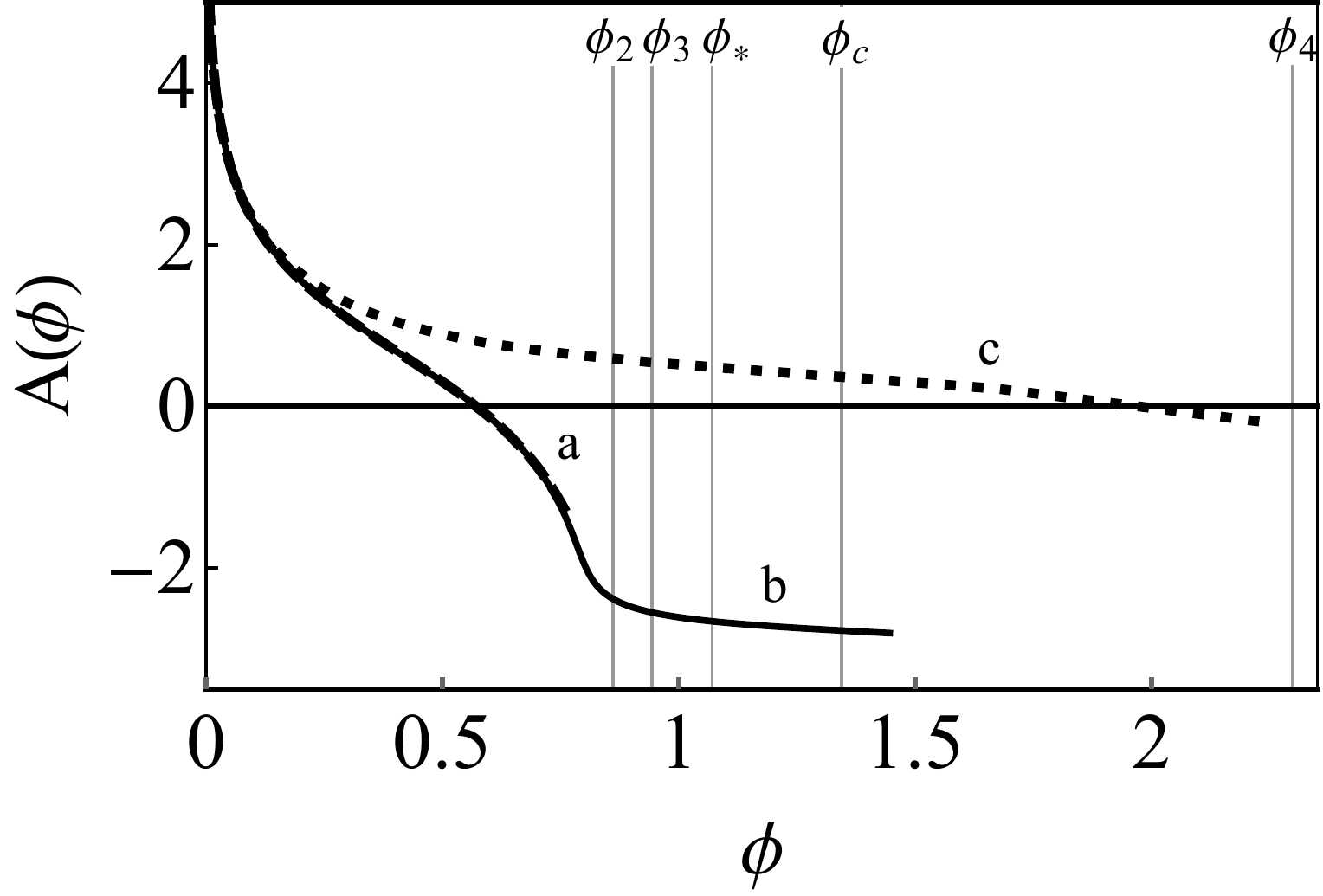}
\includegraphics[width=.49\textwidth]{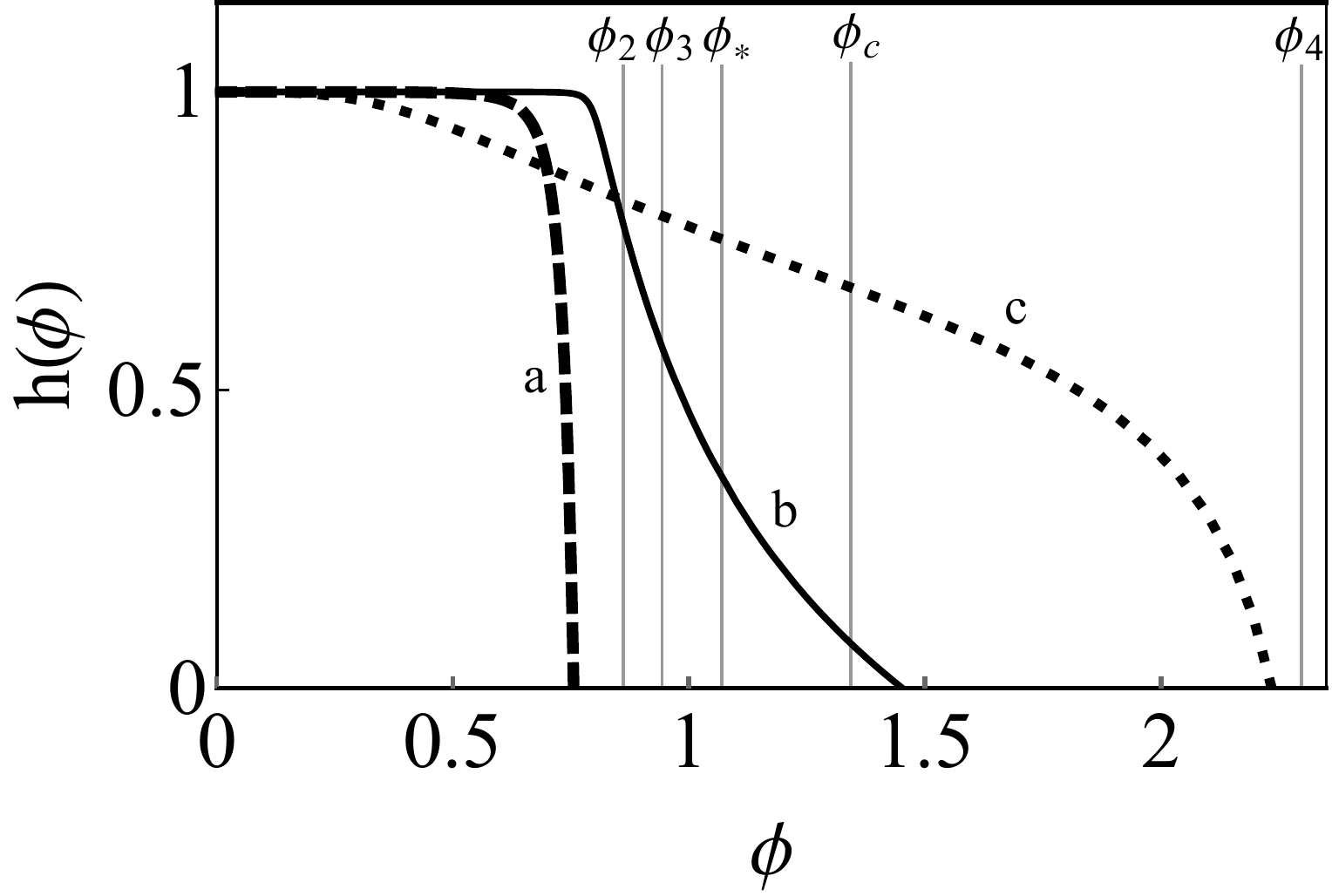}
\caption{\label{Flows_examples_FT1} \small
Some explicit examples of numerical solutions for the bulk geometry. We present three examples, one of them with $\phi_H$ in region I (case a) and two in region III (cases b and c). Each gravity solution is dual to a thermal state in the $\text{CFT}_1$.   }
\end{figure}

We will now construct the thermal states of the $\text{CFT}_1$. We parametrize the black brane solutions by $\phi_H$, defined as the value of $\phi$ at the horizon.  Let us start by heating up the non-BPS vacuum. If we start with $\phi_H$ slightly to the left of $\phi_2$, we obtain thermal states with low temperature, and as we move $\phi_H$ towards $\phi_1$, the temperature increases, and goes to infinity when $\phi_H \rightarrow \phi_1$. This corresponds to the thermal branch labeled  I in \fig{phi0_free_energy_energy}. Near the AdS points the thermodynamic functions have the expected conformal  behavior. The point labelled ``a" in \fig{phi0_free_energy_energy} indicates a thermal state on the I branch whose dual gravity solution  is displayed  in \fig{Flows_examples_FT1}.

\begin{figure}[t!!!]
\centering
\includegraphics[width=.7\textwidth]{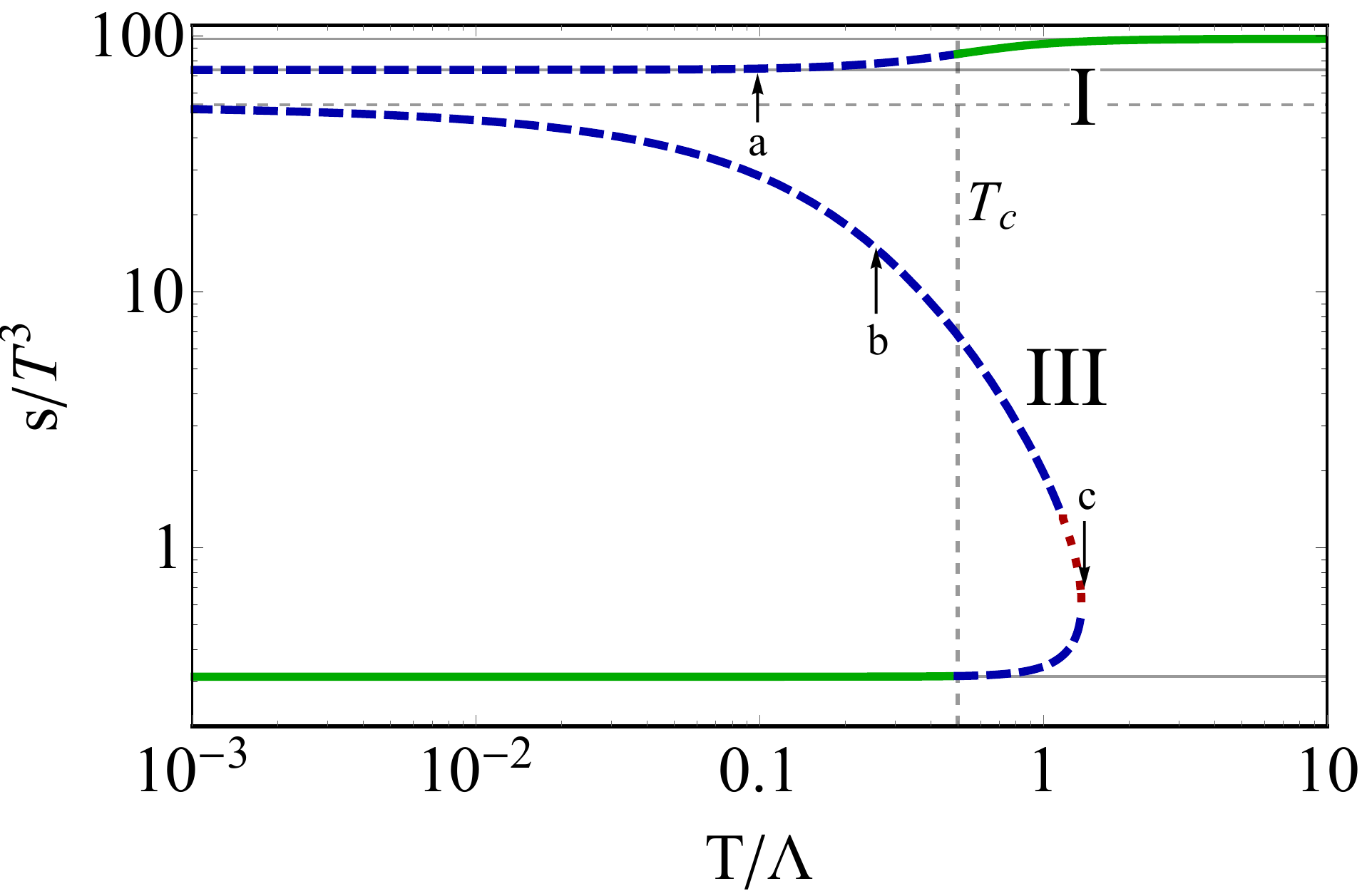}
\caption{\label{PlotProperDistanceEntropy} \small
Entropy density versus temperature of the $\text{CFT}_1$. The three  continuous horizontal  lines correspond to the entropy densities of  Schwarzschild-AdS solutions with (from top to bottom) $\phi=\phi_1$, $\phi=\phi_2$ and $\phi=\phi_4$. The dashed horizontal  line corresponds to the thermal branch of the CFT$_2$ with non-vanishing VEV.  The vertical dashed line is the phase transition temperature $T_c/\Lambda=0.499$. The labels ``a", ``b" and ``c"  indicate the thermal states dual to the gravity solutions presented in \fig{Flows_examples_FT1}.}
\end{figure}

Now let us heat up the BPS vacuum. We start with values of $\phi_H$ slightly to the left of  $\phi_4$, which correspond to low-temperature states on a second thermodynamic branch of the CFT$_1$, labelled III in  \fig{phi0_free_energy_energy}, that emanates from the BPS vaccum.  As we decrease the value of $\phi_H$ from $\phi_4$ to a certain value $\phi_c \simeq 1.3436$ located between $\phi_3$ and $\phi_4$ (see \fig{fig:potential}),  the temperature first increases, then reaches a maximum and then decreases, approaching  zero as 
$\phi_H\rightarrow\phi_c^+$. The branch III thus exhibits a maximum temperature.  Interestingly, as $\phi_H\rightarrow\phi_c^+$ the energy of the solution approaches that of the non-BPS vacuum. Note, however, that the I and III branches meet on the vertical axis with different slopes, meaning that the specific heat on each branch is different all the way down to $T=0$. This is illustrated in \fig{PlotProperDistanceEntropy}, which shows the ratio $s/T^3$, with $s$ the entropy density. We see that, although both $s$ and $T$ go to zero as $\phi_H\rightarrow\phi_c^+$, the ratio remains finite and is different on the I and III branches. 

\begin{figure}[t!!!]
\centering
\includegraphics[width=.70\textwidth]{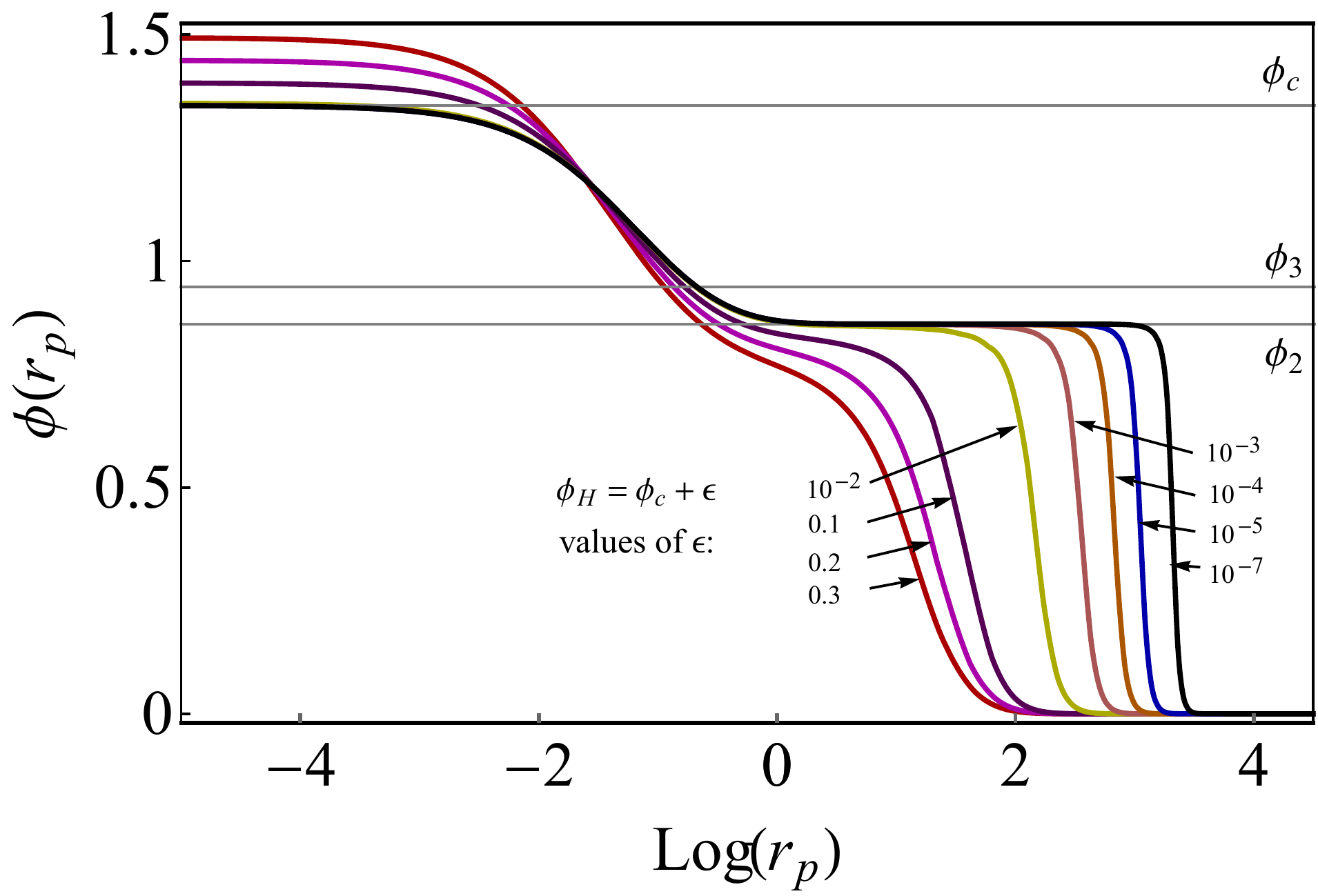}
\caption{\label{PlotProperDistance} \small
Scalar field as a function of the  proper distance $r_p$ along the holographic direction, measured from the horizon, for different solutions with $\phi_H$ approaching $\phi_c$ from the right. 
}
\end{figure}

We therefore conclude that in the limit 
$\phi_H\rightarrow\phi_c^+$ we obtain a solution which has zero temperature from the viewpoint of the $\text{CFT}_1$. This may seem in contradiction with the fact that the gravity solution seems to end  not at an extremum of the potential but at a regular horizon located at  $\phi_H=\phi_c$. Moreover, from the viewpoint of the $\text{CFT}_1$, this limiting flow has the same energy as the non-BPS vacuum described by the flow from  $\phi_1$ to  $\phi_2$.  All these features can be understood by noting that, 
as $\phi_H$ approaches $\phi_c$ from the right, the corresponding flow develops a larger and larger region in which $\phi$ is in the vicinity of 
$\phi_2$ and the metric is approximately AdS. In other words, the flow exhibits ``walking" or quasi-conformal dynamics near the fixed point described by the  $\text{CFT}_2$. This large AdS region corresponds to the plateau in \fig{PlotProperDistance}, where we plot the value of the scalar field as a function of the proper distance along the holographic direction for  several flows with $\phi_H$ close to $\phi_c$. Note that the proper distance from the plateau to the horizon approaches a finite constant as $\phi_H\rightarrow\phi_c^+$.  As the size of the plateau grows so does the relative redshift between the horizon and the UV region where $\phi \to \phi_1$. In the limit  $\phi_H\rightarrow\phi_c^+$ the flow splits  into two independent flows, one from $\phi_1$ to $\phi_2$ and one from $\phi_2$ to $\phi_c$. Since the redshift diverges in this limit, the temperature and the entropy density go to zero as seen from the  $\text{CFT}_1$. Nevertheless, we will see below that the flow from $\phi_2$ to $\phi_c$ corresponds to a state in the $\text{CFT}_2$ with non-zero temperature and entropy density.

Having introduced $\phi_c$, we can now define three regions according to the value of $\phi_H$ (see \fig{fig:potential}): 
\begin{eqnarray}
\label{def1}
&& \mbox{Region I with } \phi \in (\phi_1, \phi_2) \,, \nonumber \\[2mm]
&& \mbox{Region II with }  \phi \in ( \phi_2, \phi_{\text{c}})\,,  \\[2mm] 
&& \mbox{Region III with } \phi \in ( \phi_{\text{c}}, \phi_4) \,.\nonumber
\end{eqnarray}
What we have seen so far is that  solutions with $\phi_H$ in regions I and III are dual to thermal states of the 
$\text{CFT}_1$. As we will see shortly, solutions with $\phi_H$ outside regions I and III are not states of the $\text{CFT}_1$ because in these cases  $\phi$ does not asymptote to $\phi_1$ at large $r$.

We are now ready to discuss the thermodynamics of the $\text{CFT}_1$. From the free energy shown in \fig{phi0_free_energy_energy}(left) we conclude that there is a first order phase transition at $T_c/\Lambda \simeq0.499$. We have displayed in solid green the preferred thermodynamical states.  From the energy density shown in \fig{phi0_free_energy_energy}(right), we find a latent heat for the first order phase transition of $\epsilon/\Lambda^4 \simeq 5.2816$. The entropy density is shown in \fig{PlotProperDistanceEntropy}.
In these and in subsequent plots we show in dashed blue the locally stable but globally unstable thermal states, and in dotted red the locally unstable thermal states. Notice that in the free energy plot, for branch III, when changing from blue to red the curvature changes accordingly from convex to concave. Correspondingly, in the energy plot we see a negative  specific heat and therefore also a negative speed of sound squared. If slightly perturbed, theses homogeneous black branes would display an exponentially growing mode which would lead the system to an inhomogenous configuration, as shown in \cite{Attems:2017ezz}. 

Note that the free energy in \fig{phi0_free_energy_energy}(left) is different from the usual swallow-tail first-order phase transition that the reader may be familiar with in two related ways. First, the upper turning point lies at $T=0$, leading to a metastable dynamical supersymmetry breaking vacuum. Second, the upper branch includes both locally stable and locally unstable regions,  whereas in simpler cases these states are all locally unstable. 
We will see below that the $\text{CFT}_1$ can be continuously  deformed so that its phase diagram becomes the usual  swallow-tail first-order phase transition.

%%%%%%%%%%%%%%%%%%%%%%%%%%%%%%%%%%%%%%%%%%%%%%%%%%%%%%%%%%%%%%%%

\subsection{$\text{CFT}_2$}

Let us now consider the CFT dual to the AdS solution for which  
$\phi=\phi_2$. We are interested in flows that have this CFT as their UV fixed point. Since $\phi=\phi_2$ corresponds to a minimum of the potential the dual scalar operator has dimension $\Delta_2$ greater than four, specifically $\Delta_2 \simeq4.5069$. Turning on a source for such an irrelevant operator would destroy the UV of the theory. Therefore, we will seek flows that start at $\phi=\phi_2$ and are driven exclusively by a VEV. 

\begin{figure}[t!!!]
\centering
\includegraphics[width=.49\textwidth]{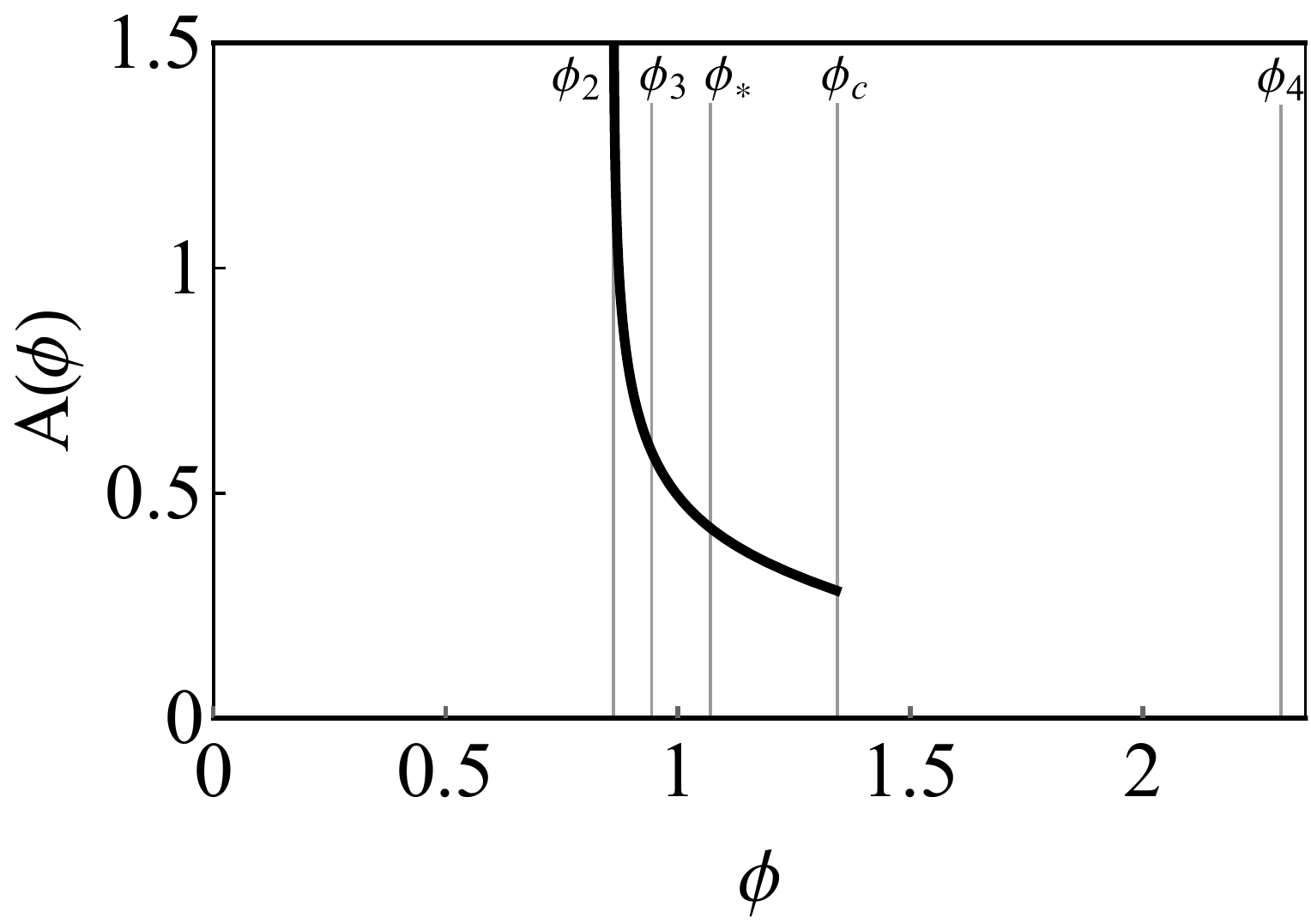}
\includegraphics[width=.49\textwidth]{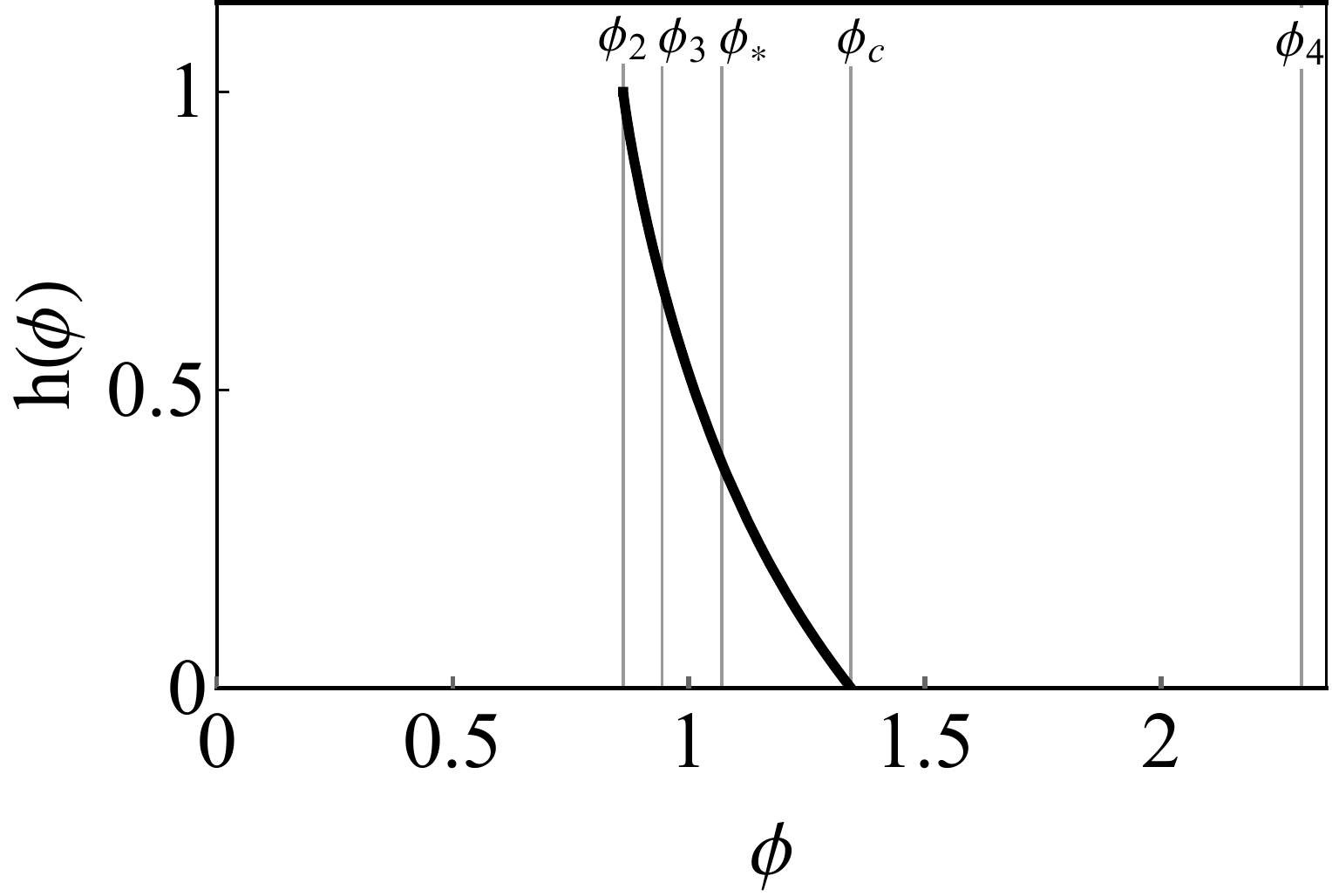}
\caption{\label{Flows_examples_FT2}  \small
Numerical solution describing the thermal state with non-zero VEV in the 
$\text{CFT}_2$. This flow may be obtained  either as the limit $\phi_H \rightarrow \phi_c^+$ or as the limit $\phi_H \rightarrow \phi_c^-$.
}
\end{figure}

The undeformed $\text{CFT}_2$ has one vacuum, given by the AdS solution with 
$\phi=\phi_2$ and $h=1$. One way to turn on a non-zero temperature is to simply replace this solution by  the corresponding Schwarzschild-AdS solution on the gravity side. In this solution the VEV of the scalar operator vanishes and therefore $T$ is the only scale. As a consequence,  any two values of  $T$ are physically equivalent to one another. Therefore one should really think of this entire branch as describing a single physically distinct thermal state. In addition to this state, the theory has a second thermal state described by the gravity solution connecting $\phi_2$ and $\phi_c$ that we found in the previous section. The functions in this solution are shown in \fig{Flows_examples_FT2}. 
Examining the behavior near $\phi_2$ we have verified that this flow is indeed triggered exclusively by a VEV of the scalar operator. Moreover, the solution possesses  a regular horizon at $\phi=\phi_c$, and therefore it describes a thermal state of the $\text{CFT}_2$. From the behaviour near the horizon and from the fall-offs  near $\phi=\phi_2$ we obtain the temperature and the entropy density for this state in units of the VEV, respectively, with the results 
\be
\frac{T}{\langle \mathcal{O} \rangle^{1/\Delta_2}}\simeq 0.5149 \sac
\frac{s}{\langle \mathcal{O} \rangle^{3/\Delta_2}} \simeq 7.333 \,. 
\ee
Note that we are referring to this solution as a thermal state and not as a branch of states. The reason is that any two states are physically distinguished only by the dimensionless ratio $T/\langle \mathcal{O} \rangle^{1/\Delta_2}$, which as we have seen is uniquely fixed by the solution.

We therefore conclude that, at any non-zero $T$, the $\text{CFT}_2$ has two physically distinct thermal states, one with $T/\langle \mathcal{O} \rangle^{1/\Delta_2}=0$ and one with $T/\langle \mathcal{O} \rangle^{1/\Delta_2}\simeq0.5149$. Comparing their free energies we have verified that the thermodynamically preferred state is the one with vanishing VEV.  Interestingly, the state with vanishing VEV can be viewed as the IR limit of  the thermal branch of the $\text{CFT}_1$ whose gravity flow ends at $\phi_2$, whereas the state with non-vanishing VEV can be viewed as the IR limit of  the thermal branch of the $\text{CFT}_1$ whose gravity flow ends at $\phi_c$. This is illustrated, for example, by their entropy-to-temperature ratios. Indeed, we see in  \fig{PlotProperDistanceEntropy} that the zero-temperature limit of $s/T^3$ for the upper thermal branch of the 
$\text{CFT}_1$ agrees precisely with the value of this ratio in the zero-VEV thermal state of the  $\text{CFT}_2$, represented by the continuous horizontal line corresponding to $\phi=\phi_2$. Similarly, the zero-temperature limit of the lower thermal branch of the $\text{CFT}_1$ that ends at the non-BPS state agrees precisely with the  value of this ratio in the non-zero-VEV thermal state of the  $\text{CFT}_2$, represented by the dashed horizontal line.

%%%%%%%%%%%%%%%%%%%%%%%%%%%%%%%%%%%%%%%%%%%%%%%%%%%%%%%%%%%%%%%%

\subsection{CFT$_3$}

Recall that we have chosen the parameters of our superpotential so that the dimension of the scalar operator in the CFT$_3$ is $\Delta_3=3$. The flows that we will describe  in this section are therefore characterised  by a source 
$\Lambda$ for this  operator. Since the potential is not symmetric around $\phi_3$, flows with a positive source are physically distinct from flows with a negative source, and we will therefore consider each case separately.  In terms of gravity solutions, this means that the scalar field 
$\phi$ will asymptote to $\phi_3$ from the left for a negative source and from the right for a positive source. 
 
\begin{figure}[t!!!]
\centering
\includegraphics[width=.49\textwidth]{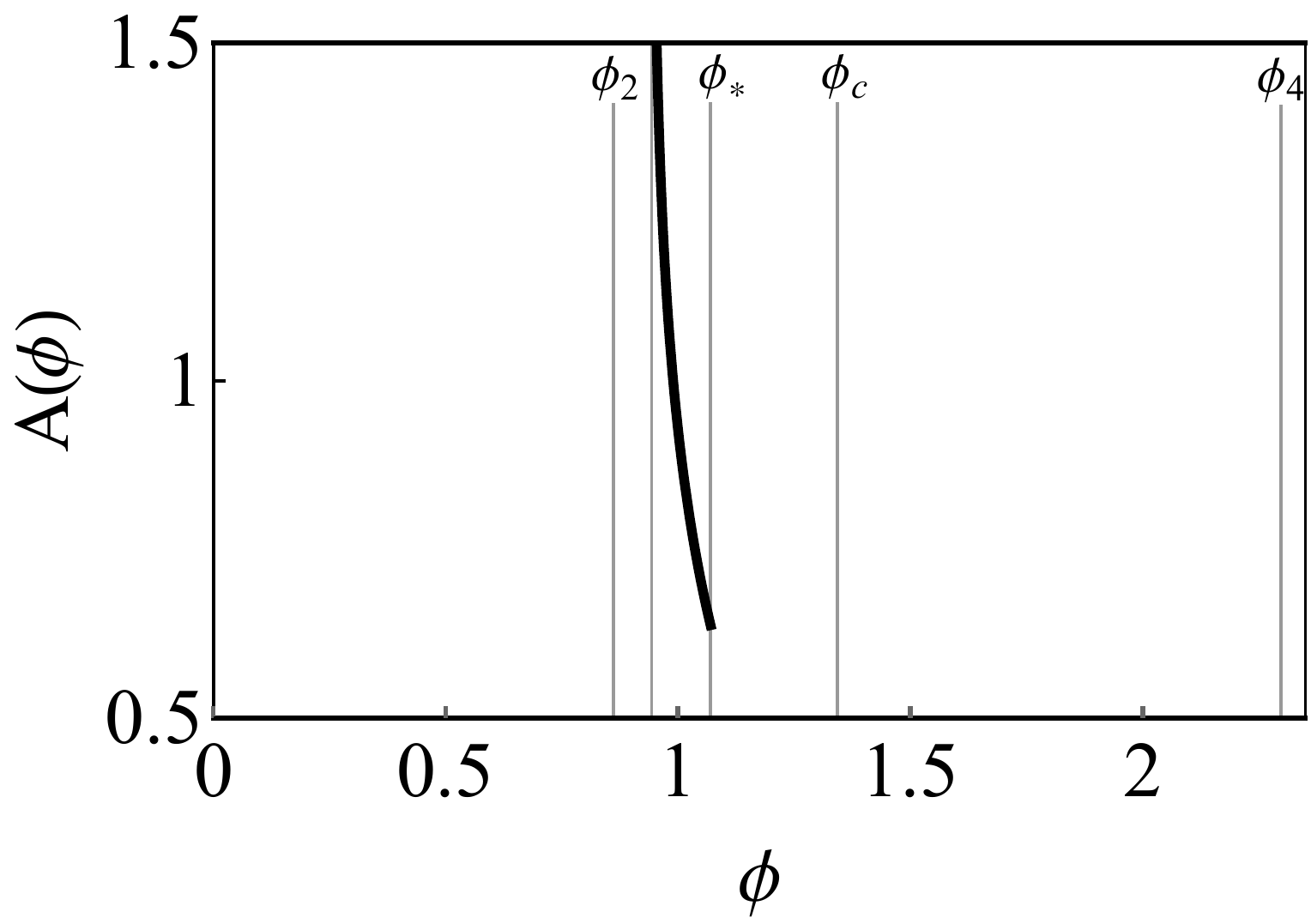}
\includegraphics[width=.49\textwidth]{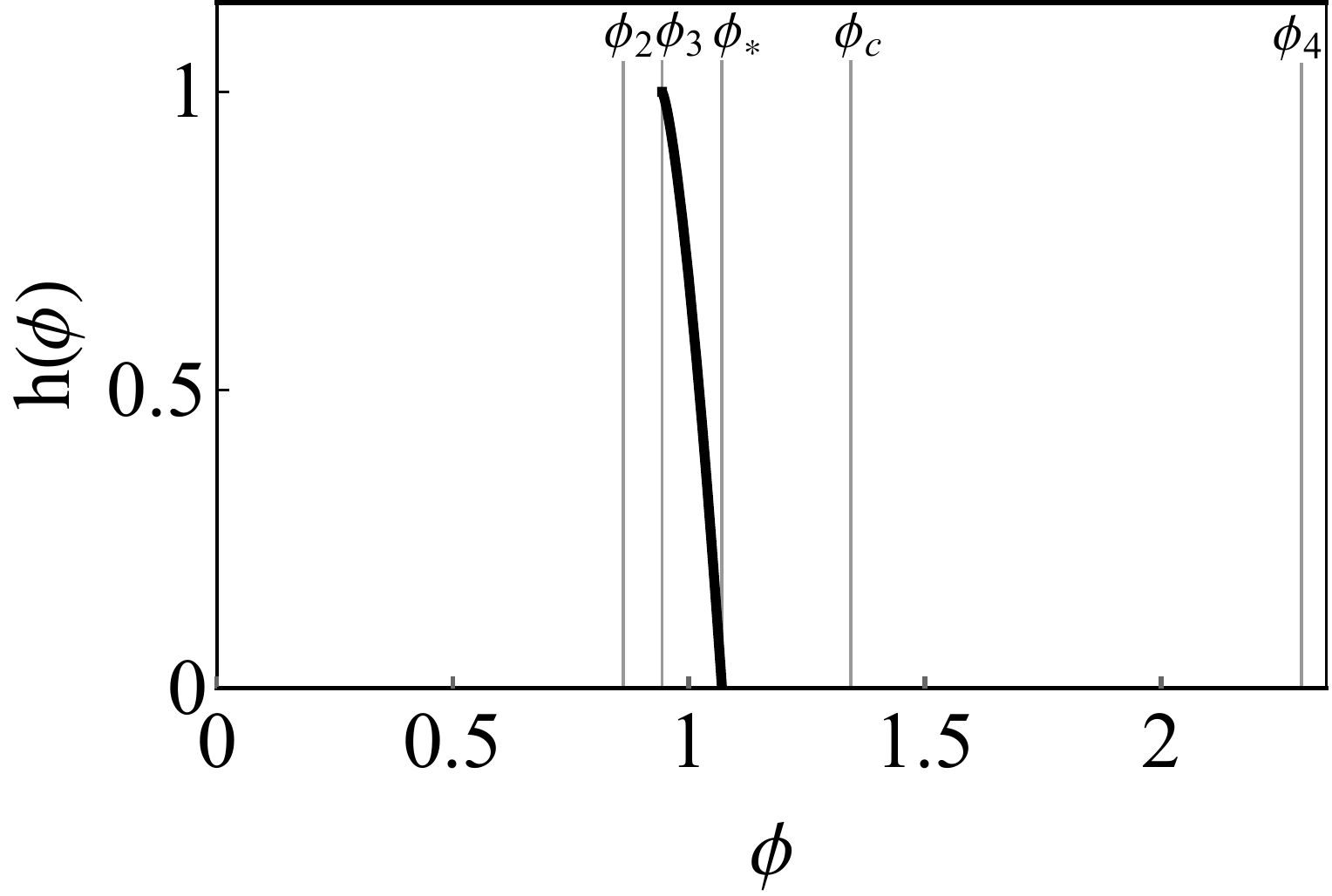}
\caption{\label{Flows_examples_FT3_VEVnonzero}  \small
Numerical solution describing the thermal state with non-zero VEV in the 
$\text{CFT}_3$. This flow is obtained for $\phi_H=\phi_*$.
}
\end{figure}

We start by studying the theory with vanishing source $\Lambda=0$. The vacuum is given by the AdS solution with $\phi=\phi_3$ and $h=1$.  If we consider the zero-source theory at non-zero temperature, we find two physically thermal states, in analogy with the case of the CFT$_2$. The first thermal state is described by the AdS-Schwarzschild solution with $\phi=\phi_3$. The second thermal state corresponds to a gravity solution in which the scalar field starts in the UV at 
$\phi=\phi_3$ and increases monotonically until the flow ends at a regular horizon located at a point given by $\phi_* \simeq 1.0701$ (see \fig{fig:potential}). This flow is shown in \fig{Flows_examples_FT3_VEVnonzero}.
 From the viewpoint of the CFT$_3$ with zero source, these two states are distinguished by the fact that the first one has vanishing VEV whereas the second one has 
\be
\frac{T}{\langle \mathcal{O} \rangle^{1/\Delta_3}} \simeq 0.6649 \sac 
\frac{s}{\langle \mathcal{O} \rangle^{3/\Delta_3}} \simeq 21.26 \,.  
\ee
Comparing their free energies we have verified that the thermodynamically preferred state is the one with vanishing VEV. 

We will now deform the CFT$_3$ by turning on a source for the scalar operator. Since this operator is relevant, the UV of the theory will still be described by the undeformed CFT$_3$. We therefore expect that the two thermal branches of the sourceless theory will be reflected in the 
high-temperature physics of the deformed theory. 

In order to classify the different gravity solutions with a $\phi(r)$ that asymptotes to $\phi_3$ for large $r$ it is useful to divide region II into three subregions: 
\begin{eqnarray}
\label{def2}
&& \mbox{Region IIa with } \phi \in (\phi_2, \phi_3) \,, \nonumber \\[2mm]
&& \mbox{Region IIb with }  \phi \in ( \phi_3, \phi_*) \,, \\[2mm] 
&& \mbox{Region IIc with } \phi \in ( \phi_*, \phi_c) \,. \nonumber
\end{eqnarray}
Solutions with $\phi_H$ in IIb asymptote to $\phi_3$ from the right, so they correspond to thermal states of the CFT$_3$ deformed by a positive source. Solutions with $\phi_H$ in IIa and IIc asymptote to $\phi_3$ from the left, so they correspond to thermal states of the CFT$_3$ deformed by a negative source.  The difference between these two cases is that solutions with 
$\phi_H$ in IIa are monotonic, whereas solutions with $\phi_H$ in IIc exhibit a  ``bounce", namely, they are non-monotonic in $\phi$: they first decrease and go below $\phi_3$, then they have a bounce, and finally asymptote to 
$\phi_3$ from the left.  The point $\phi_*$ is defined as the point where the gravity solutions start to develop a bounce. From the field theory viewpoint, it is a limiting point where the source changes sign from positive to negative, so at exactly $\phi_H=\phi_*$ we have a sourceless solution, as explained above. 

\subsubsection{Negative source}
\begin{figure}[t!!!]
\includegraphics[width=.49\textwidth]{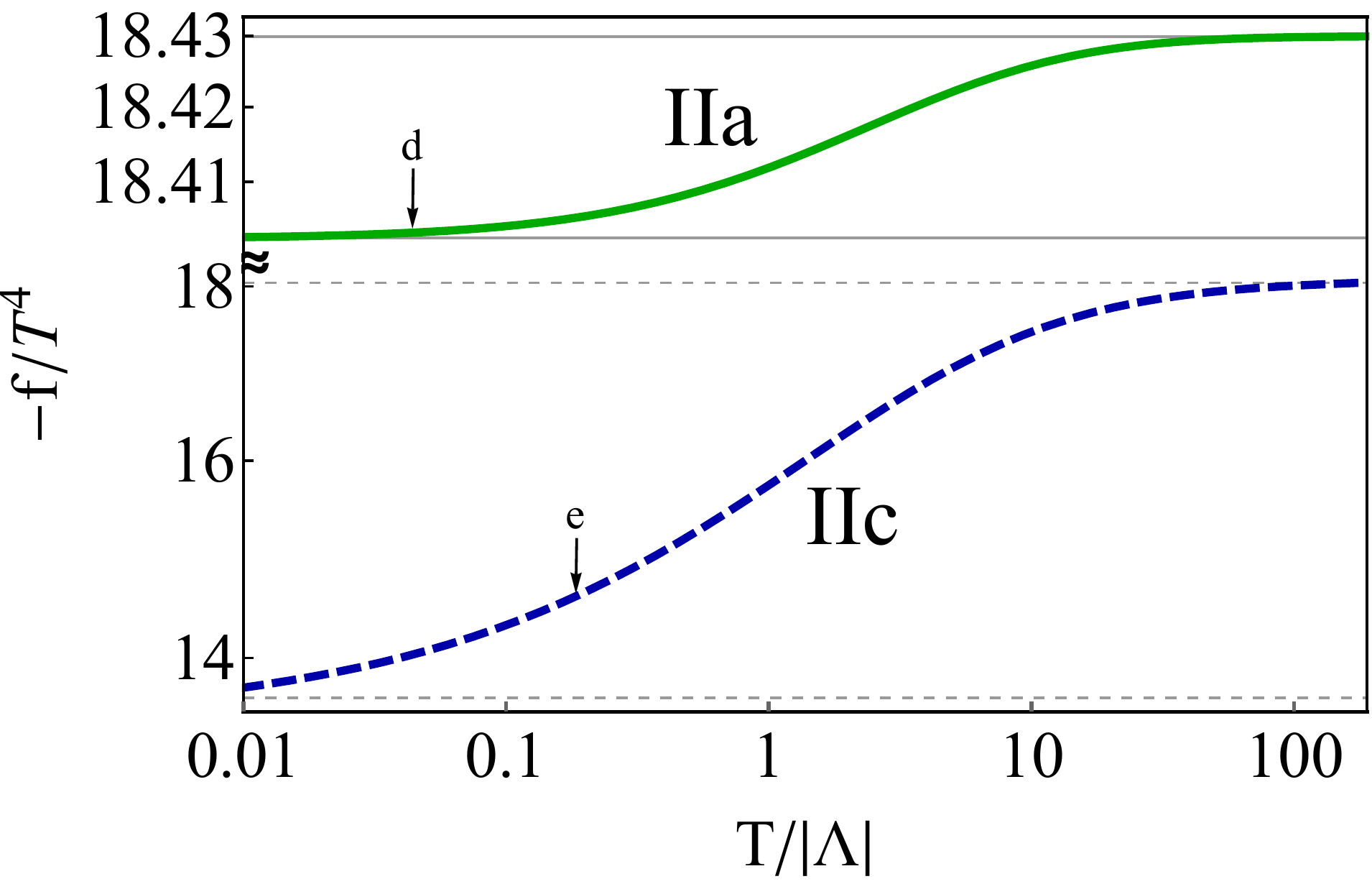}
\includegraphics[width=.49\textwidth]{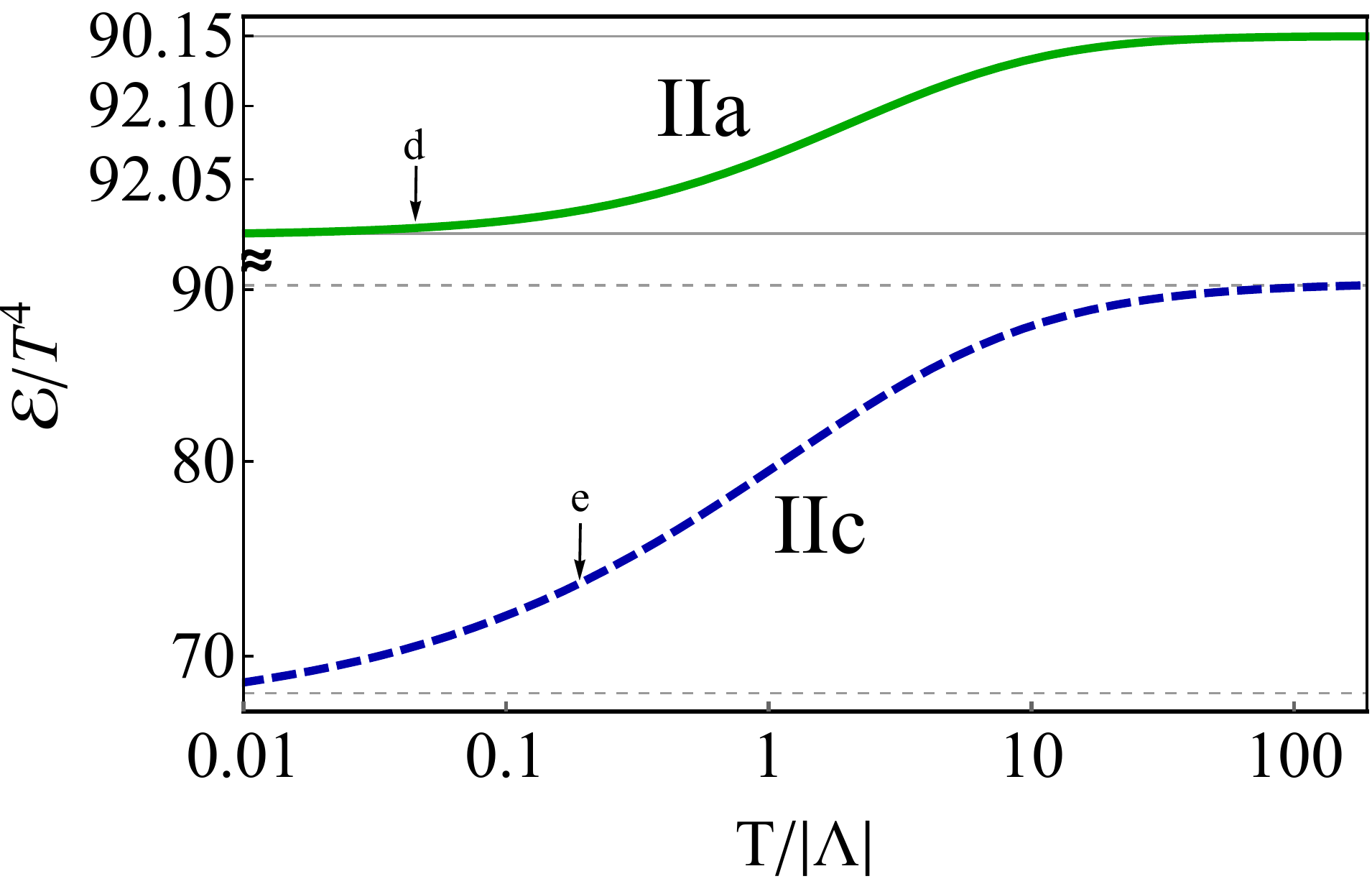}
\caption{\label{FT3_NegSource_free_energy_energy} \small
Free energy density (left) and energy density (right) versus temperature of the CFT$_3$ with negative source $\Lambda$. 
The two continuous horizontal lines correspond to the thermodynamic quantities of Schwarzschild-AdS solutions with (from top to bottom) $\phi=\phi_3$ and  
$\phi=\phi_2$.  
The dashed horizontal lines correspond to the thermal branches  (from top to bottom) of the CFT$_3$ and of the CFT$_2$ with non-vanishing VEV.
The locally stable but globally unstable branch is shown in dashed blue, whereas the thermodynamically preferred branch is shown in solid green.}
\end{figure}

\begin{figure}[t!!!]
\centering
\includegraphics[width=.49\textwidth]{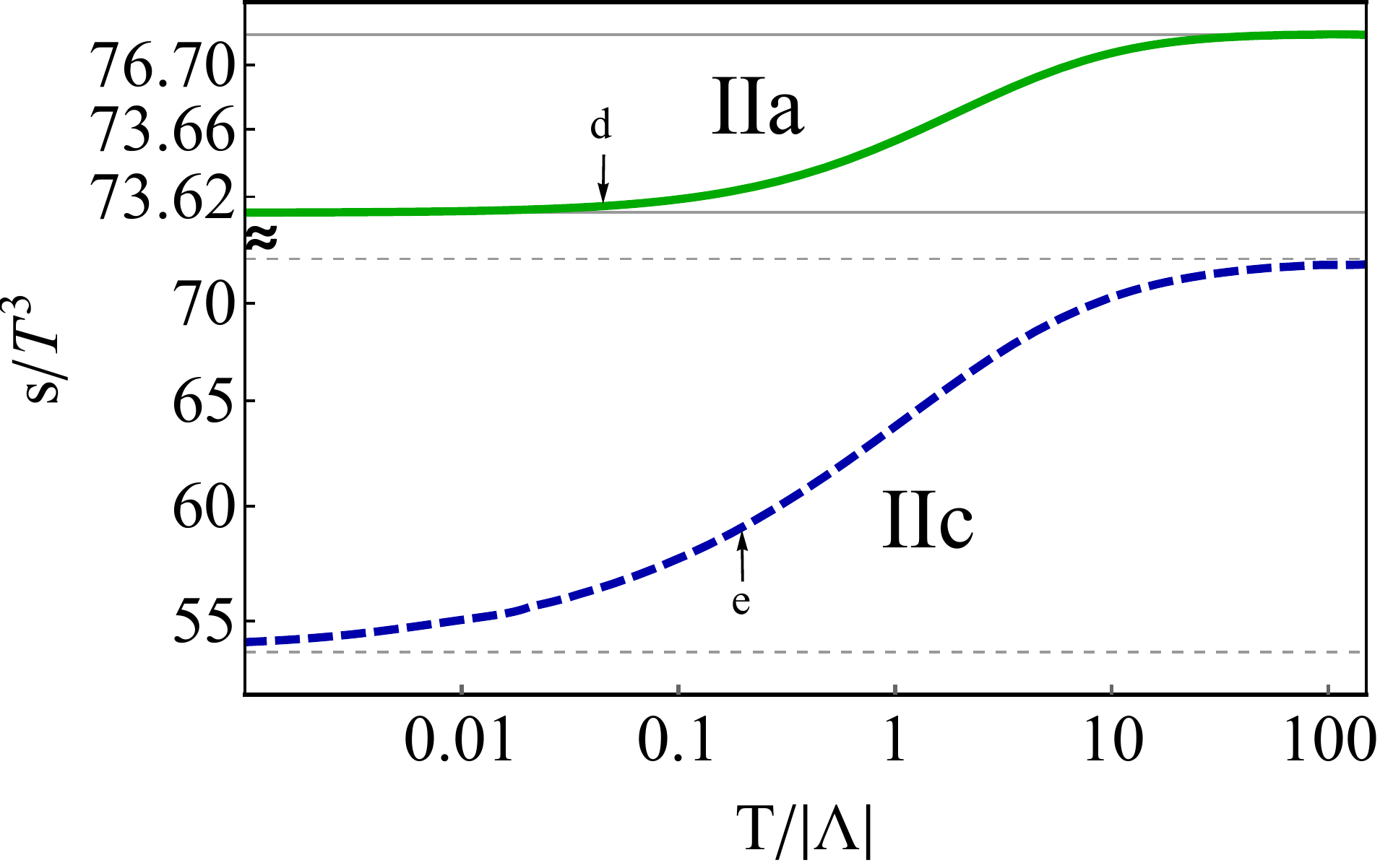}
\includegraphics[width=.49\textwidth]{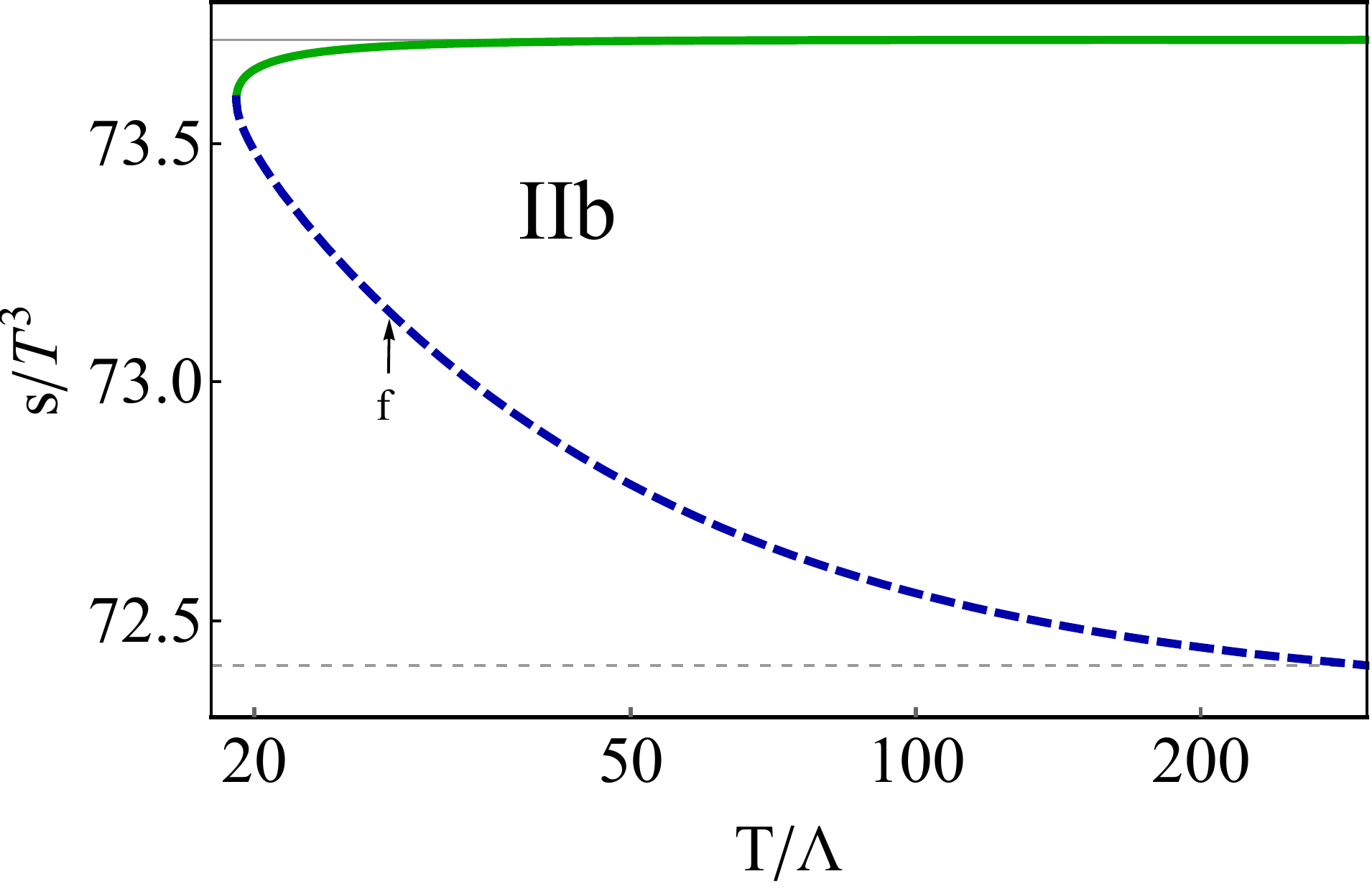} 
\caption{\label{soverT3_vs_T_FT3_PosSource} \small
Entropy density versus temperature of the CFT$_3$ with negative  (left) and positive (right) source $\Lambda$. The horizontal lines are as in \fig{FT3_NegSource_free_energy_energy}.
In dashed blue locally stable but globally unstable and in solid green thermodynamically preferred. Regions of the potential IIa, IIb, IIc, map to the thermodynamic branches here. The labels d, e and f  are the values for the thermal states dual to the gravity solutions presented in \fig{Flows_examples_FT3}.}
\end{figure}

\begin{figure}[tbhp]
\centering
\includegraphics[width=.49\textwidth]{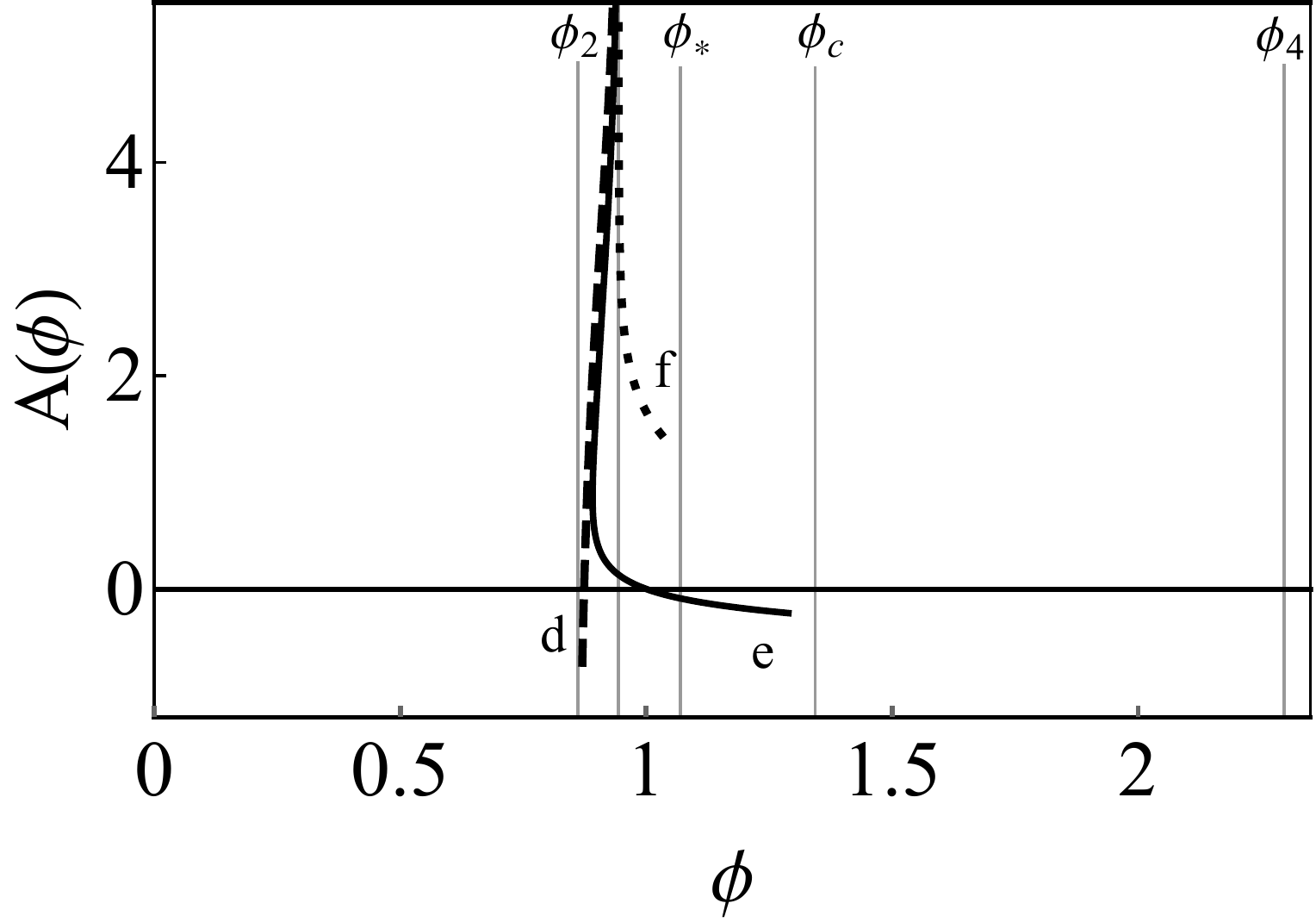}
\includegraphics[width=.495\textwidth]{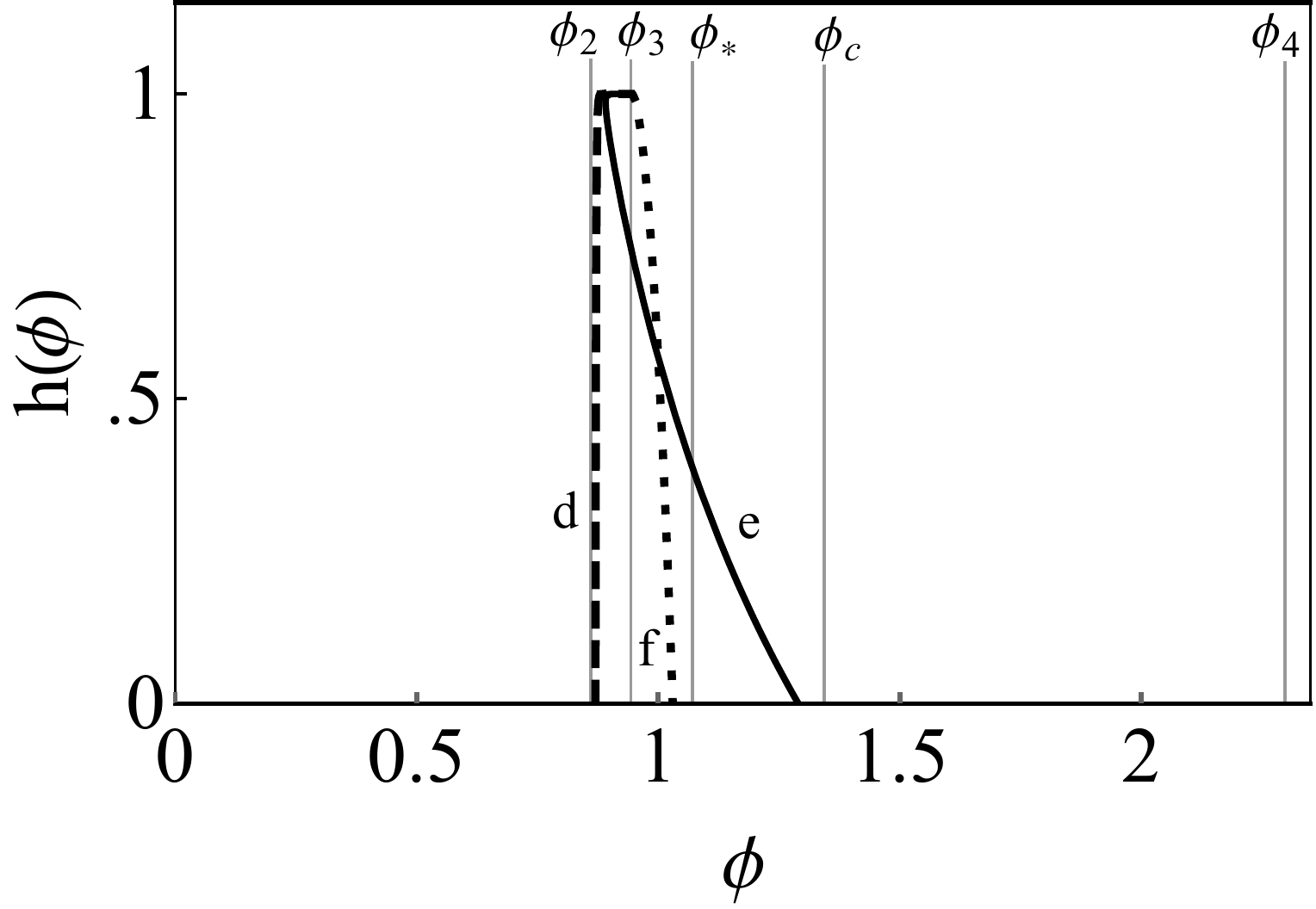}
\caption{\label{Flows_examples_FT3} \small
Several numerical solutions of the bulk geometry with values of $\phi_H$ in region II. Solution d is an example with $\phi_H$ in region IIa, solution e with $\phi_H$ in region IIc and f $\phi_H$ in region IIb. 
  }
\end{figure}

Thermal states of the CFT$_3$ deformed by a negative source correspond to gravitational solutions with $\phi_H$ in regions IIa and IIc. The vacuum of this theory is given by a flow from $\phi_3$ to $\phi_2$ with $h=1$, which corresponds to a smooth interpolation between two AdS solutions. Notice that this flow is not a solution to the BPS equations obtained from the superpotential (\ref{superpotential}). Now we can heat up this vacuum solution, and as we move $\phi_H$ from $\phi_2$ to $\phi_3$ we reconstruct the thermodynamic branch labeled as IIa in \fig{FT3_NegSource_free_energy_energy} and \fig{soverT3_vs_T_FT3_PosSource}(left). Values of  $\phi_H$ near $\phi_2$  correspond to low temperatures, and the thermodynamics corresponds to the thermodynamics of the AdS-Schwarzschild for the CFT$_2$. Values of 
$\phi_H$ near $\phi_3$  correspond to high temperatures, and the thermodynamics corresponds to the thermodynamics of the  AdS-Schwarzschild solution for the CFT$_2$. Therefore, this IIa branch smoothly interpolates between the two zero-source, zero-VEV thermal branches of the  CFT$_2$ and the CFT$_3$, respectively. For an example of a numerical solution with $\phi_H$ in IIa see \fig{Flows_examples_FT3}, solution d. 

Consider now the solutions with $\phi_H$ in IIc. Recall that these solutions exhibit a bounce. If $\phi_H$ is close to $\phi_*$ from the right, the bounce is small, and this solution corresponds to large temperatures and the thermodynamics is given by the thermodynamics of the non-zero-VEV thermal branch of the CFT$_3$. As we increase $\phi_H$, the bounce becomes larger, and when $\phi_H$ gets close to $\phi_c$ the turning point gets closer to $\phi_2$. In the limit $\phi_H \rightarrow \phi_c^-$, the flow decouples into  two flows, one  from $\phi_3$ to $\phi_2$ and one from $\phi_2$ to $\phi_c$. This decoupling is analogous to the decupling in the limit $\phi_H \rightarrow \phi_c^+$ that we  studied in the CFT$_1$. Again, there is an infinite redshift and the dual field theory state has zero temperature from the viewpoint of the CFT$_3$. In particular, this implies that there are two Lorentz-invariant states, to which we will refer as  non-degenerate vacua. In summary, the states  with $\phi_H$ in IIc give rise to the thermodynamic branch  labelled as IIc in \fig{FT3_NegSource_free_energy_energy} and \fig{soverT3_vs_T_FT3_PosSource}(left). This branch interpolates smoothly between the sourceless CFT$_3$ branch with non-vanishing VEV at large temperatures and the sourceless CFT$_2$ branch with non-vanishing VEV at low temperatures. For an example of a numerical solution with $\phi_H$ in IIc see \fig{Flows_examples_FT3} , solution e.

We thus conclude that the CFT$_3$ deformed with a negative source possesses two thermal branches, labelled IIa and IIc in \fig{FT3_NegSource_free_energy_energy} and \fig{soverT3_vs_T_FT3_PosSource}(left). One of them connects the AdS-Schwarzschild solutions at the endpoints, and the other connects the non-zero-VEV branches. Nevertheless, the thermodynamics of the CFT$_3$ does not exhibit any phase transitions. The thermodynamically preferred states are those of IIa, and the states IIc are only metastable. There are no locally unstable thermal states in the CFT$_3$. The high-temperature physics reflects the two thermal branches of the zero-source theory and the low-temperature physics reflects the two thermal branches of the zero-source CFT$_2$.

\subsubsection{Positive source}
\begin{figure}[t!!!]
\includegraphics[width=.49\textwidth]{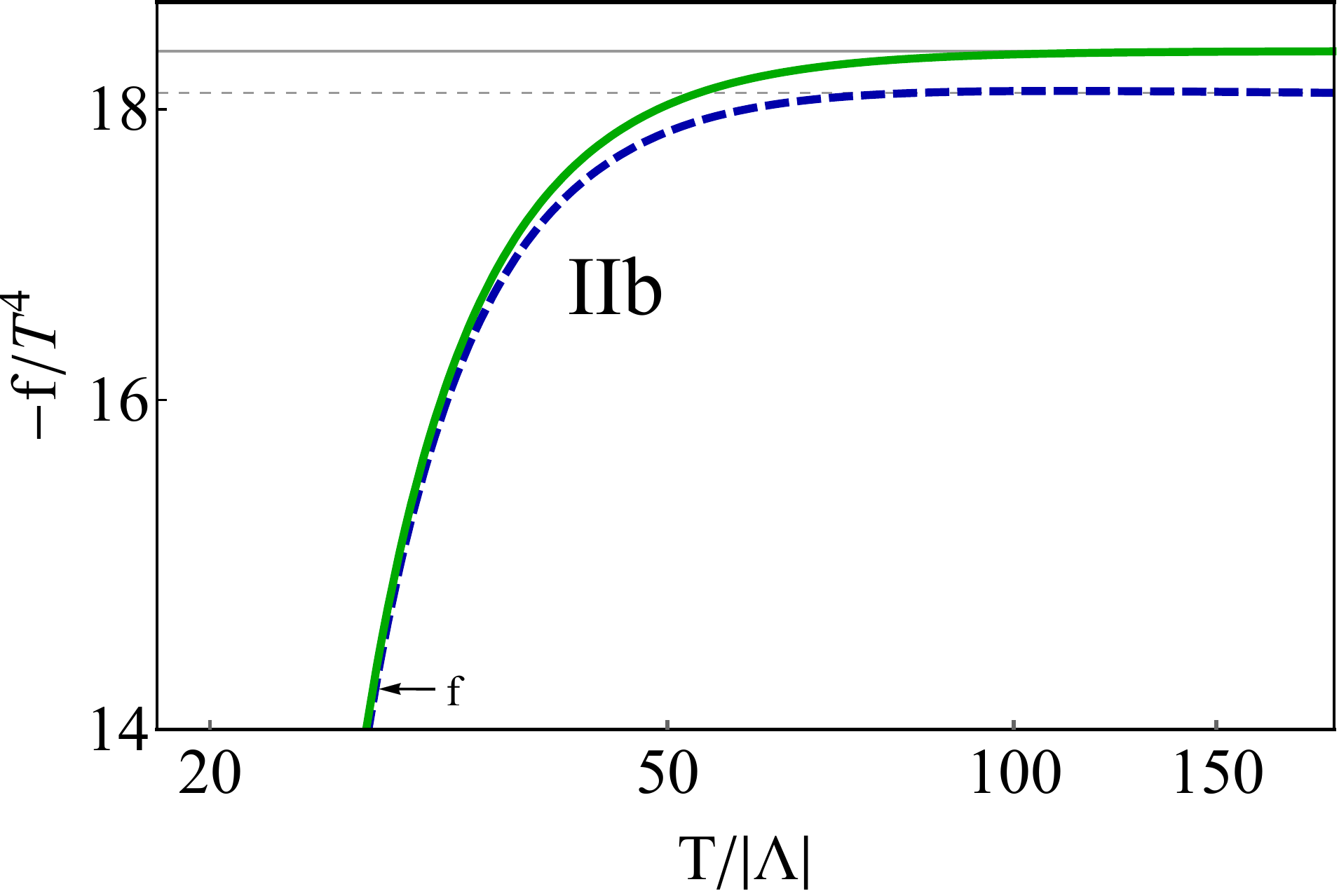}
\includegraphics[width=.49\textwidth]{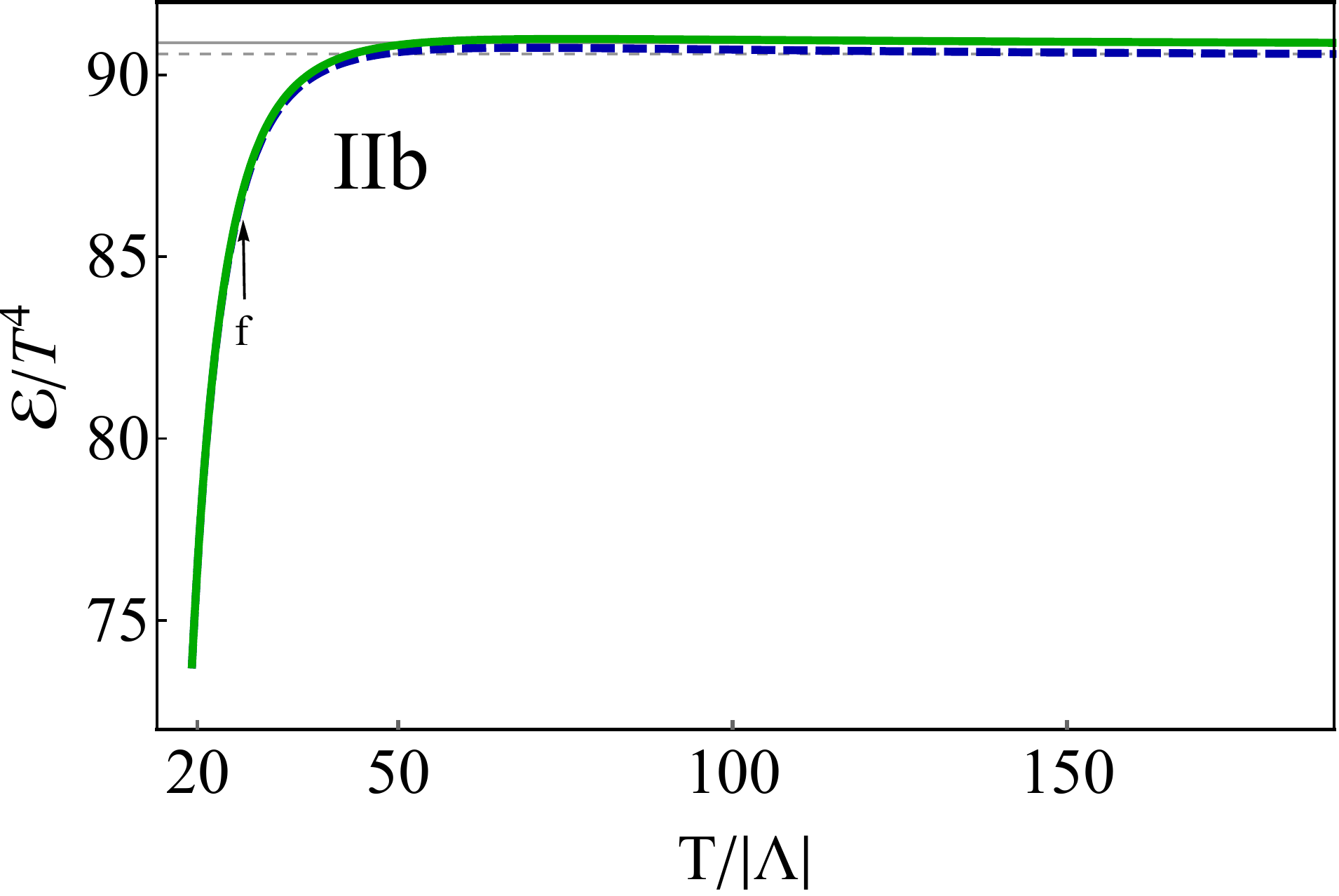}
\caption{\label{FT3_PosSource_free_energy_energy} \small
Free energy (left) and energy density (right) versus temperature of the CFT$_3$ with a positive source  $\Lambda$. Stable (metastable) states are shown in solid green (dashed blue). }
\end{figure}

Let us now consider the CFT$_3$ deformed with a positive source. On the gravity side this corresponds to solutions with $\phi_H$ in IIb that asymptote to $\phi_3$ from the right. These solutions do not bounce, and the scalar field increases monotonically from $\phi_3$ to $\phi_H$. For an example of one of these numerical solutions see \fig{Flows_examples_FT3}, solution f. If we start with $\phi_H$ slightly to the right of  $\phi_3$ we obtain thermal states with high temperature, and  the thermodynamics is approximately  that of the undeformed CFT$_3$. If we now increase the value of  $\phi_H$ towards $\phi_*$, the temperature first decreases, then reaches a minimum value $T_{\text{min}}/\Lambda \simeq 19.12$,  and then  increases, eventually diverging as $\phi \to \phi_*^-$. This behaviour is clearly illustrated in  \fig{FT3_PosSource_free_energy_energy} and \fig{soverT3_vs_T_FT3_PosSource}(right). Near this value, the thermodynamics approaches the thermal branch of the CFT$_3$ with zero source and non-zero VEV.  The existence of a minimal temperature means that our analysis is unable to identify a candidate ground state of the CFT$_3$ deformed with a positive source. A possible reason is that such a source destabilizes the zero-temperature theory, and that only in the presence of a sufficiently large temperature the thermodynamic ensemble  becomes well defined. 

Note that the  limit  $T/\Lambda \rightarrow \infty$ can be thought  of as the zero-source limit. Indeed, in this limit we  recover precisely the sourceless cases, since the thermodynamics asymptotes to the  AdS-Schwarzschid thermal branch  and to the branch with non vanishing VEV. This is somehow the inverse situation of  that for  the III branch of the CFT$_1$: In that case there is a maximum temperature and the thermodynamics  asymptotes to the two sourceless thermal branches for  $T/\Lambda \rightarrow 0$.

\subsection{CFT$_4$}
We close this section with some brief comments on this case. This theory  has only one vacuum, the AdS solution with $\phi=\phi_4$ and $h=1$. It has only one thermal branch,  the AdS-Schwarzschild solution with  $\phi=\phi_4$. These features are reflected in the IR physics of the CFT$_1$. Since $\phi_4$ is a minimum of the potential the dimension of the scalar operator is larger than 4. Turning on a source for such an irrelevant operator would therefore destroy the UV and we have not found any flows that start at $\phi_4$ and are triggered by a VEV.

%%%%%%%%%%%%%%%%%%%%%%%%%%%%%%%%%%%%%%%%%%%%%%%%%%%%%%%%%%%%%%%%%%%%%%%
%%%%%%%%%%%%%%%%%%%%%%%%%%%%%%%%%%%%%%%%%%%%%%%%%%%%%%%%%%%%%%%%%%%%%%%
%%%%%%%%%%%%%%%%%%%%%%%%%%%%%%%%%%%%%%%%%%%%%%%%%%%%%%%%%%%%%%%%%%%%%%%
%%%%%%%%%%%%%%%%%%%%%%%%%%%%%%%%%%%%%%%%%%%%%%%%%%%%%%%%%%%%%%%%%%%%%%%

\section{Related theories}

So far we have studied  the theory (\ref{eq:action}) with the superpotential (\ref{superpotential}) and $\phi_M=0.5797$. In this section we study the theory for other values of $\phi_M$. These new examples will shed light on  some  aspects of the thermodynamics of exotic RG flows that we have encountered above.

%For example, the connection of the thermodynamics of CFT1 presented in, with the usual swallow tail thermodynamics.

%%%%%%%%%%%%%%%%%%%%
%%%%%%%%%%%%%%%%%%%%
%%%%%%%%%%%%%%%%%%%%
%%%%%%%%%%%%%%%%%%%%%

\subsection{Connection with the usual  swallow-tail}

In this subsection we establish a connection between the thermodynamics of the CFT$_1$ studied in Section \ref{subsectionCFT1} and the thermodynamics of a theory with the usual swallow-tail first-order phase transition. We obtain one theory as a continuous deformation of the other. 

Specifically, we smoothly modify the superpotential (\ref{superpotential}) by continuously increasing the value of the parameter $\phi_M$ from $\phi_M= 0.5797$ to $\phi_M=0.9$, while keeping $\phi_Q=10$ constant. For the initial value $\phi_M= 0.5797$, the thermodynamics of the CFT$_1$ is displayed in \fig{phi0_free_energy_energy}. 
\begin{figure}[t!!!]
\centering
\includegraphics[width=.485\textwidth]{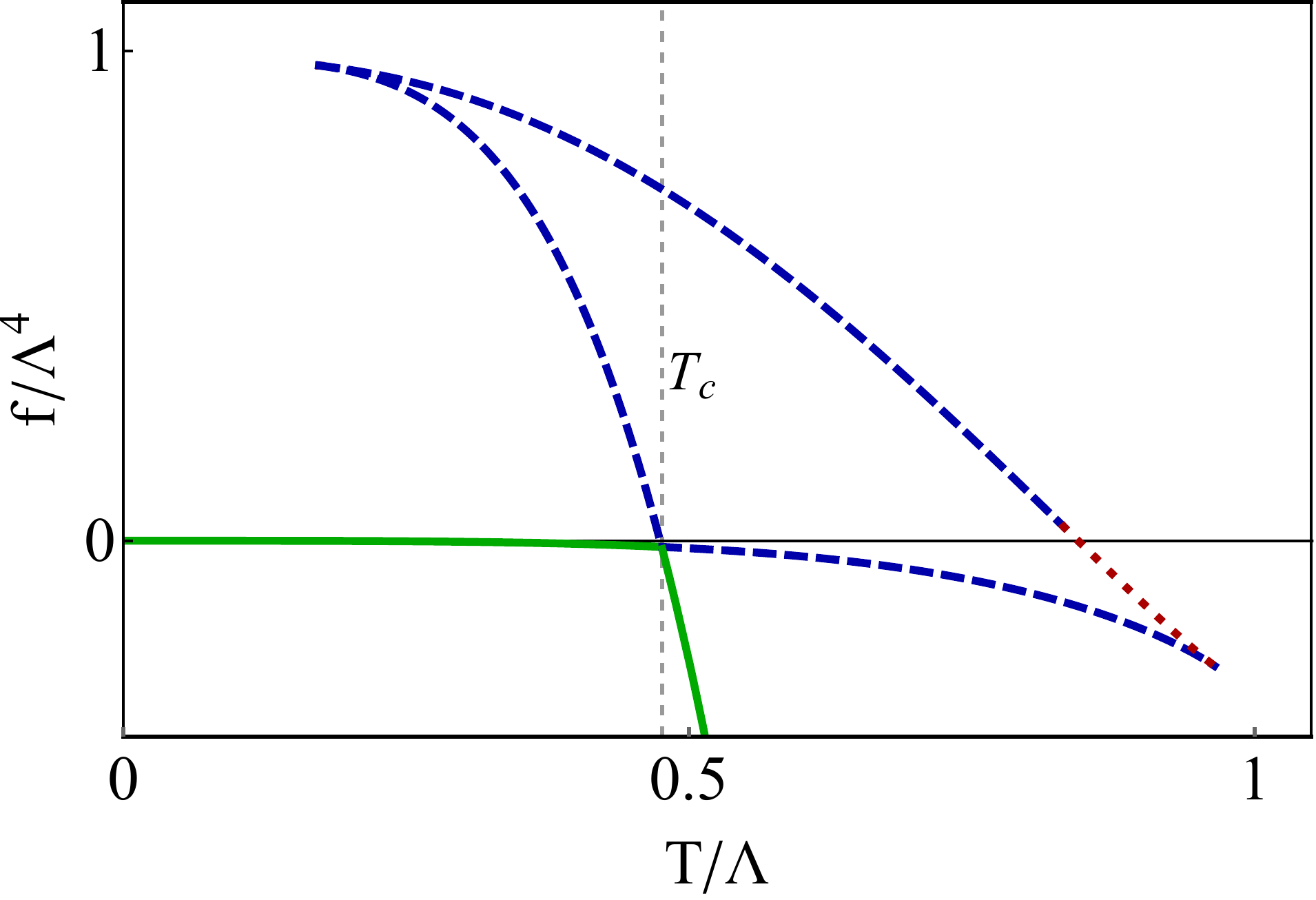}
\includegraphics[width=.49\textwidth]{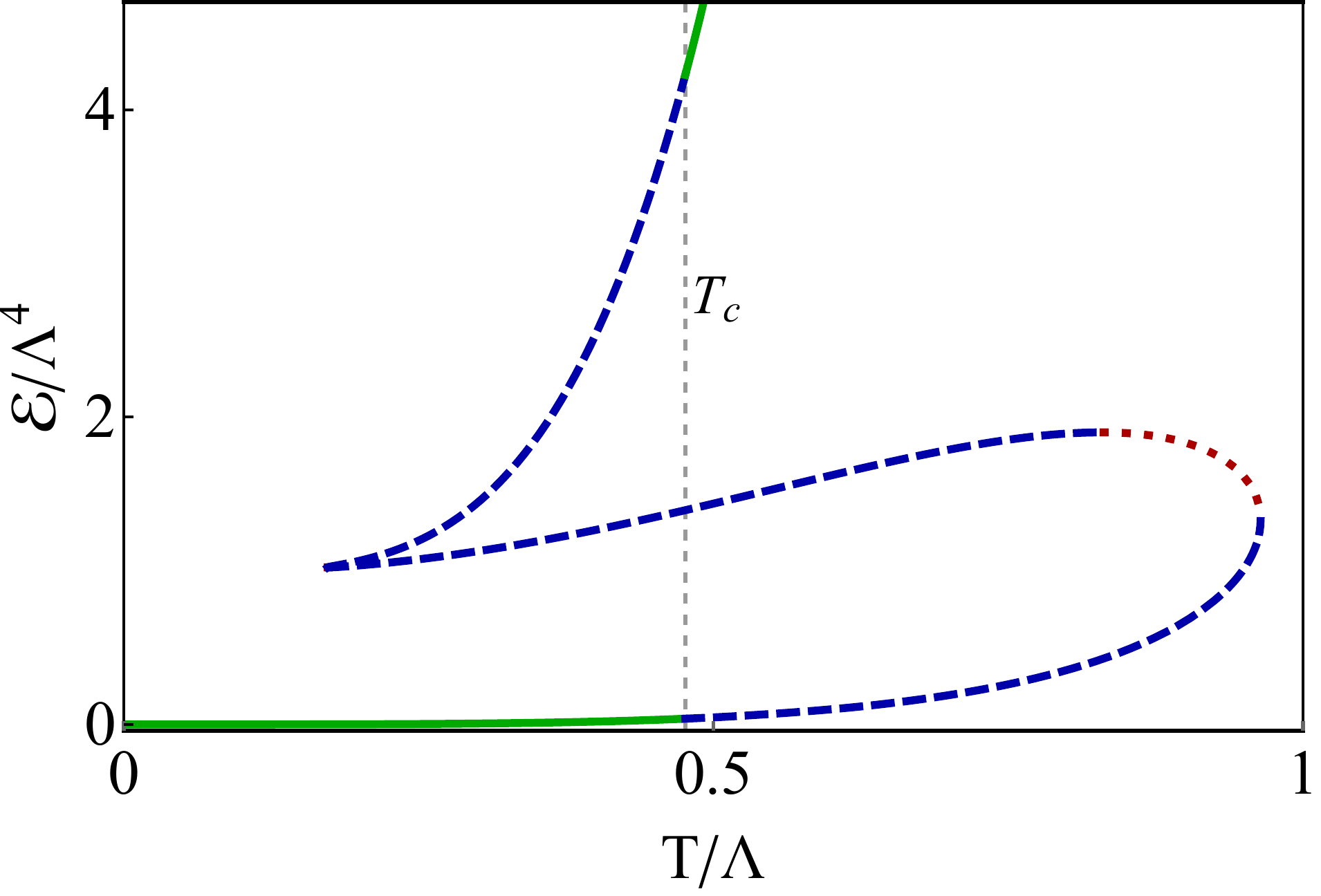}
\includegraphics[width=.499\textwidth]{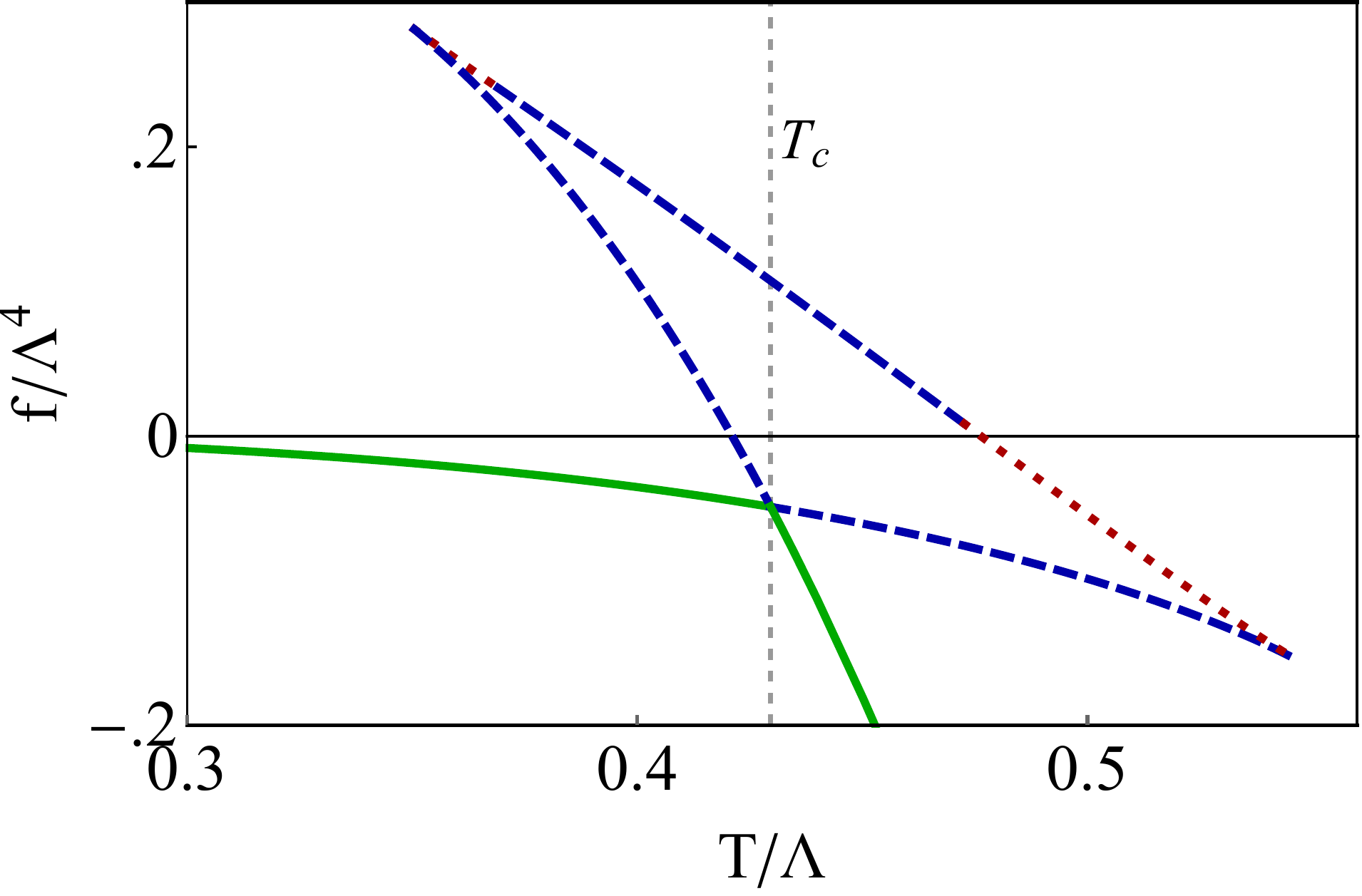}
\includegraphics[width=.485\textwidth]{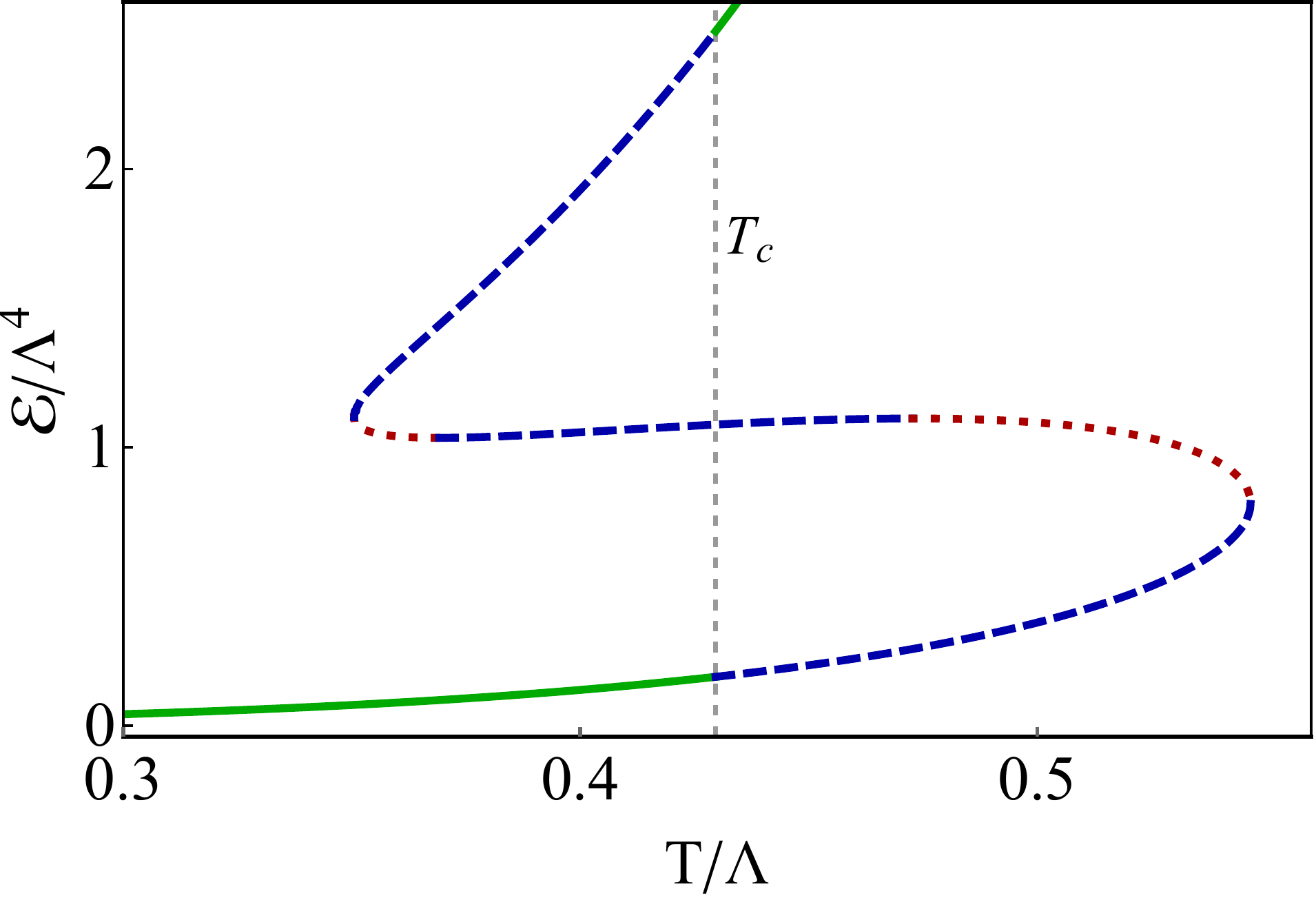}
\includegraphics[width=.499\textwidth]{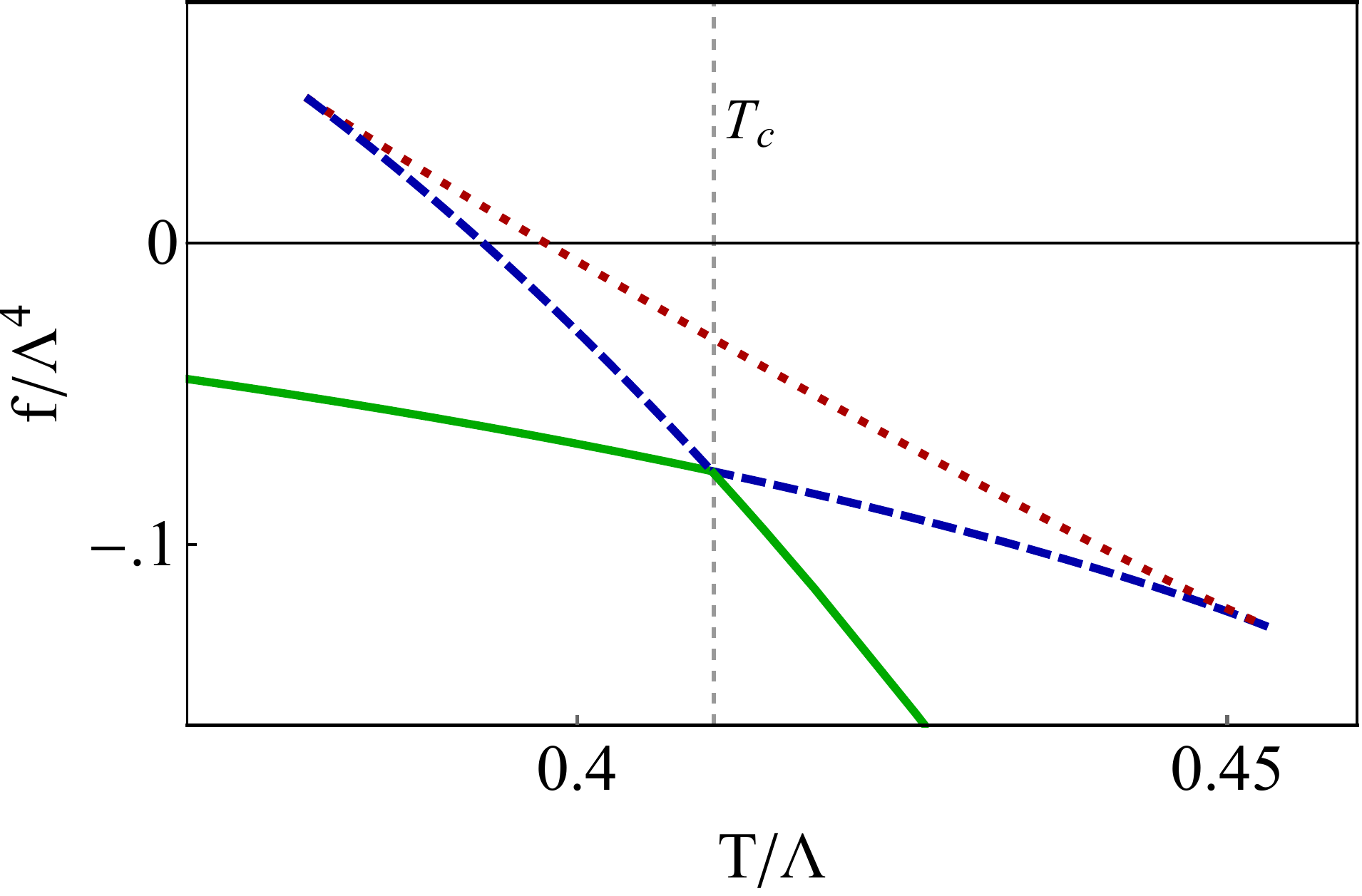}
\includegraphics[width=.485\textwidth]{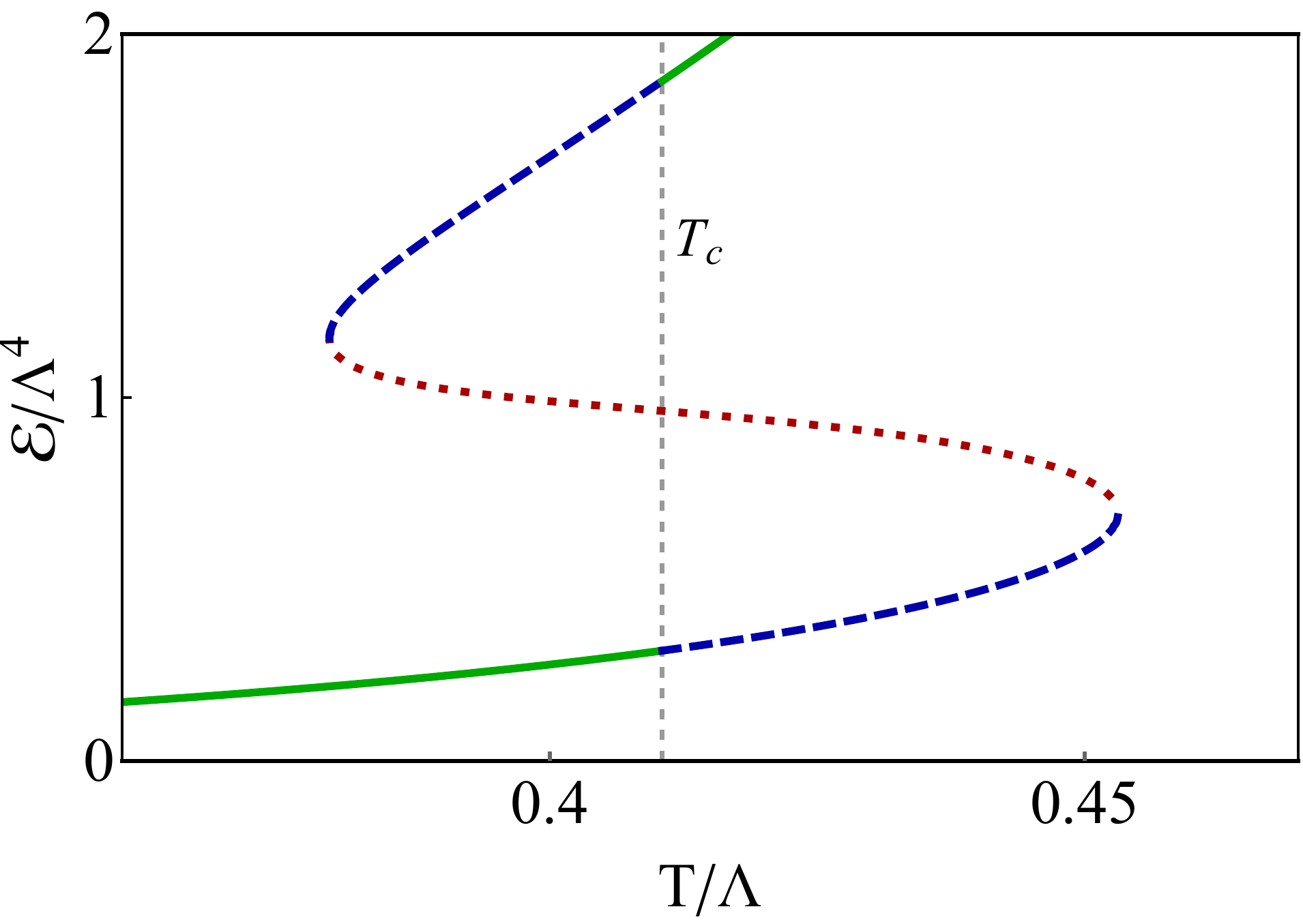}
\caption{\label{connection_with_swallow_tail} \small
Free energy (left) and energy density (right) versus temperature for (from top to bottom)  $\phi_M=0.64$, $\phi_M=0.8$, $\phi_M=0.9$.}
\end{figure}
As we increase the value of $\phi_M$ the two non-supersymmetric extrema  approach each other, and for $\phi_M = 0.5808$ they merge into an inflection point (we study this particular case in the next section). For larger values of $\phi_M$ there are no non-supersymmetric extrema of the potential in between the two extrema of the superpotential. This implies that the non-supersymmeric vacuum of the CFT$_1$ does not exist. As a consequence, the upper thermal branch of this theory does not touch the $T=0$ axis any more. This is illustrated in  \fig{connection_with_swallow_tail}(top), which corresponds to $\phi_M=0.64$. In this case the turning point of the energy density is no longer a kink, and there are locally unstable phases in the two turning regions, separated by a region of locally stable states. This is more clearly seen in \fig{connection_with_swallow_tail}(center), which corresponds to 
$\phi_M=0.8$. If we keep increasing $\phi_M$ further we eventually recover the usual swallow-tail situation in which there is a single locally unstable region, as shown in \fig{connection_with_swallow_tail}(bottom), which corresponds to $\phi_M=0.9$.

For all these values of $\phi_M$ the CFT$_1$ possesses a first-order phase transition at the temperature $T_c$ indicated by a vertical line in \fig{connection_with_swallow_tail}. If we increase $\phi_M$ even further, then the latent heat of the first order phase transition becomes smaller and smaller, until it vanishes for  $\phi_M\simeq 1.2$. At this point the theory possesses a second-order phase transition. For larger values  of $\phi_M$ the transition becomes a smooth crossover.

%%%%%%%%%%%%%%%%%%%%
%%%%%%%%%%%%%%%%%%%%
%%%%%%%%%%%%%%%%%%%%
%%%%%%%%%%%%%%%%%%%%%

\subsection{Inflection point}
We will now study the thermodynamics of the theory with $\phi_M\simeq 0.5808$. For this value the two non-supersymmetric extrema merge with one another giving rise to an inflection point at $\phi_2=\phi_3 \simeq 0.9015$. Similarly, the two points $\phi_*$ and $\phi_c$ also merge with one another at 
$\phi_*=\phi_c \simeq 1.219$. As a result, the original regions IIa and IIc disappear, and we are left only with region IIb. Regions I and III remain qualitatively similar. The non-supersymmetric vacuum  for the CFT$_1$ still exists as the flow from $\phi_1$ to $\phi_2=\phi_3$. Thus the thermodynamics of the  CFT$_1$ remains qualitatively unchanged with respect to that of Section \ref{subsectionCFT1} (\fig{phi0_free_energy_energy}). Notice that the skipping flows, those with $\phi_H$ in region III, skip only one CFT in this case.

Consider now the CFT defined by the inflection point. At leading order, the scalar operator dual to the scalar field is a marginal operator because the first and second derivatives of the potential vanish. However, beyond leading order the operator becomes marginally irrelevant if the source is negative, and marginally relevant if the source is positive \cite{Kiritsis:2016kog}. For vanishing source, this CFT has only one vacuum given by the AdS solution with $\phi=\phi_2=\phi_3$, and it has two thermal branches, the AdS-Schwarzschild solution, and a flow from 
$\phi_2=\phi_3$ to $\phi_*=\phi_c$. This flow corresponds to a solution with non vanishing VEV for the scalar operator. Turning on a negative source would destroy the UV since in this case the operator is marginally irrelevant. If instead the source is positive, then the operator is marginally relevant, and all the thermal solutions starting in region IIb are thermal states of this CFT. Thus this theory is the analog of the  CFT$_3$ deformed by a positive source
 that we studied above. 
 
It is interesting to ask what happens if we now increase $\phi_M$ slightly above 0.5808. In this case the inflection point disappears and so does the CFT at that point. Region IIb disappears and regions I and III merge.
What is then the fate of solutions which used to have  $\phi_H$ in region IIb? The inflection point disappears, but there is a region of the potential which has near-zero derivative, so it gives rise to a near-AdS solution. Thus, the corresponding solutions ``stay for a long time'' in this quasi-conformal  region before reaching  $\phi_1$. The thermodynamic branches no longer touch the $T=0$ axis but they get very close to it due to the large redshift associated to the near-AdS region. Thus the states that used to be thermal states of the CFT at $\phi_2=\phi_3$ now become thermal states of the CFT$_1$. 

%%%%%%%%%%%%%%%%%%%%
%%%%%%%%%%%%%%%%%%%%
%%%%%%%%%%%%%%%%%%%%
%%%%%%%%%%%%%%%%%%%%%

\subsection{Potential with a positive maximum}
\label{positive_potential_section}
\begin{figure}[t!!!]
\centering
\includegraphics[width=.49\textwidth]{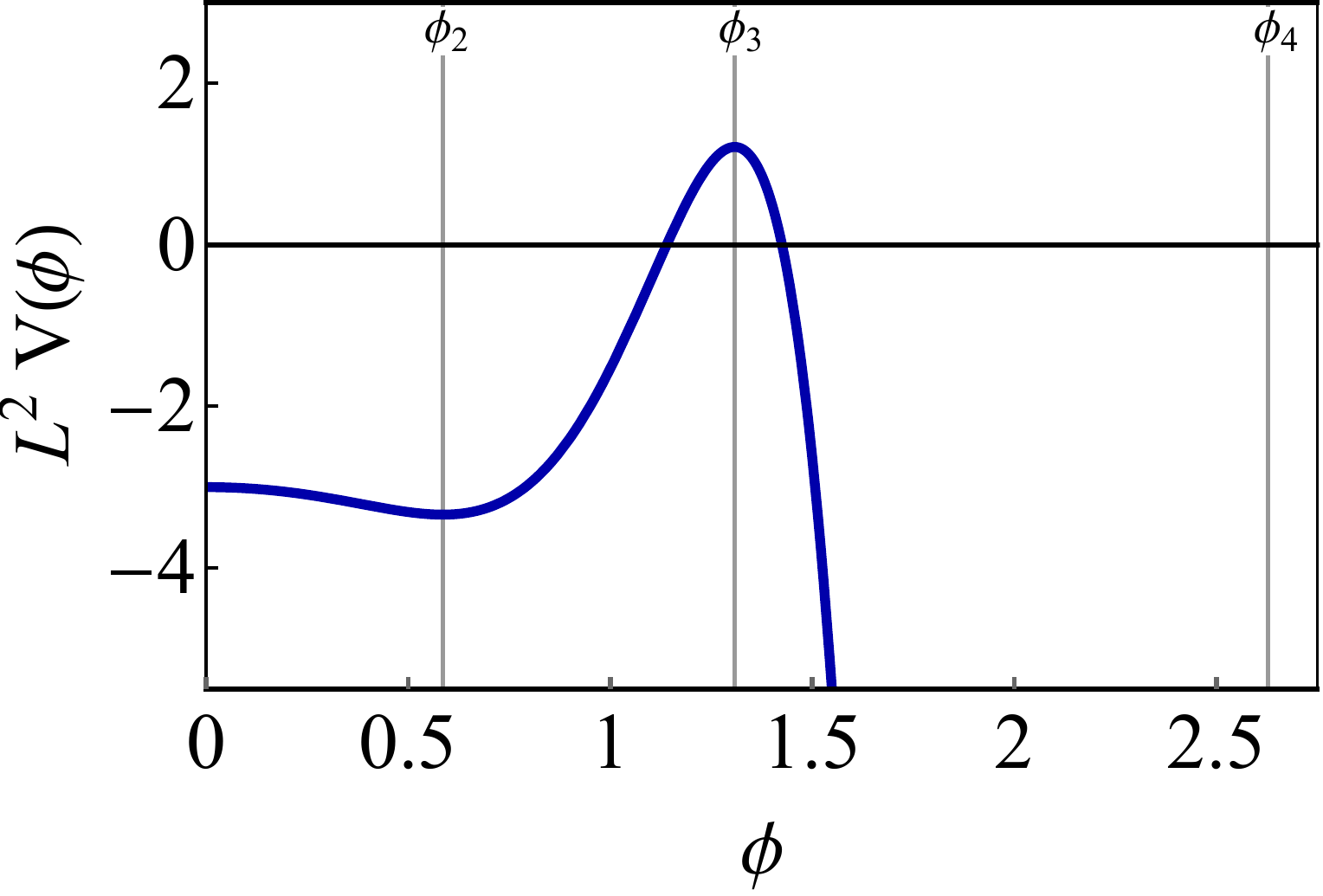}
\includegraphics[width=.49\textwidth]{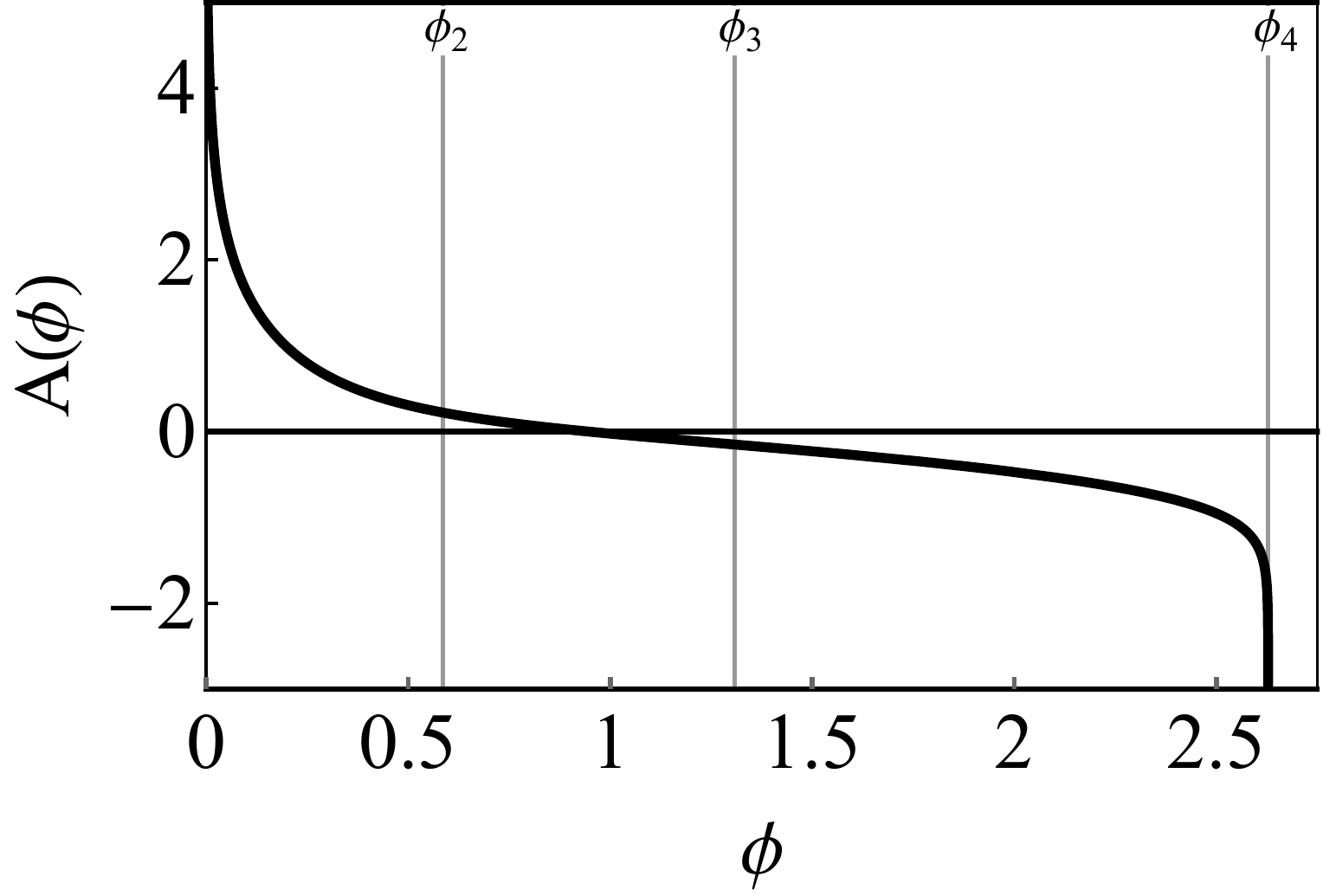}
\caption{\label{positive_potential_plots}  \small
Left: Region of the potential with a positive maximum. Right: skipping flow solution for zero temperature ($h=1$).}
\end{figure}

It is interesting that by varying the parameter $\phi_M$ one can continuously  deform the potential in such a way that the non-supersymmetric maximum at $\phi_3$ lies in the positive region. For illustration purposes, we choose $\phi_M=0.5$, for which the potential is shown in \fig{positive_potential_plots}(left). The existence of a globally defined superpotential guarantees that the supersymmetric, skipping flow that connects $\phi_1$ to $\phi_4$ still exists. In fact, we have verified that the entire thermodynamics of the CFT$_1$ is qualitatively identical to that discussed above. 

Constant-scalar solutions with $\phi=\phi_3$ still exist, but the spacetime metric is now dS instead of AdS. It would be interesting to find flows that connect an AdS extremum to this dS solution.

%%%%%%%%%%%%%%%%%%%%%%%%%%%%%%%%%%%%%%%%%%%%%%%%%%%%%%%%%%%%%%%%%%%%%%%
%%%%%%%%%%%%%%%%%%%%%%%%%%%%%%%%%%%%%%%%%%%%%%%%%%%%%%%%%%%%%%%%%%%%%%%
%%%%%%%%%%%%%%%%%%%%%%%%%%%%%%%%%%%%%%%%%%%%%%%%%%%%%%%%%%%%%%%%%%%%%%%
%%%%%%%%%%%%%%%%%%%%%%%%%%%%%%%%%%%%%%%%%%%%%%%%%%%%%%%%%%%%%%%%%%%%%%%

\section{Interpolating between exotic RG flows at zero-temperature}
\label{interpol}

So far we have focussed on the study of thermal states described by exotic RG flows. In this section we consider the zero-temperature case.  Exotic RG flows at $T=0$ can be classified as:\footnote{The fourth case is not considered an exotic flow in the terminology of  \cite{Kiritsis:2016kog}.} 
\begin{eqnarray}
& & \bullet ~ \text{Bouncing flows} \nonumber \\
& & \bullet ~ \text{Skipping flows} \nonumber \\
& & \bullet ~ \text{Irrelevant VEV flows}\nonumber  \\
& & \bullet ~ \text{Relevant VEV flows} \nonumber
\label{different_RG_flows}
\end{eqnarray}
In this section we review these types of flows and present a potential that allows us to interpolate continuously between  all of them. 

For presentation purposes, we consider a concrete example of a potential with a free paremeter $g$:
\begin{equation}
L^2 V(\phi)=-3-\frac{3}{8} \phi^2+\phi^4- g \ \phi^6 + 0.23 \ \phi^8 ~,
\label{potential_exotic_RG_flows}
\end{equation}
where $L$ is a lengh scale. By varying the parameter $g$, we find the different cases of exotic RG flows. We only consider zero temperature solutions ($h=1$). For the range of $g$ that we study, $g \in (0.84, 0.99)$, the qualitative features of the potential are always the same. In particular, it has two maxima  that we denote by $\phi_1$ and $\phi_3$, and two minima that we denote by $\phi_2$ and $\phi_4$, as in the rest  of the paper. Nevertheless, we will see that the structure of the flows does change  qualitatively as $g$ varies within the range above. To illustrate this we focus on flows that end on the minimum at $\phi_4$ from the left, which corresponds to turning on a negative source for the irrelevant operator at 
$\phi_4$. The absolute value of the source is irrelevant since  it is the only scale at the CFT dual to $\phi_4$. We then adopt an IR viewpoint in the sense that we start at $\phi_4$ as an IR fixed point and ask what the possible UV origin of the flow is for different values of $g$. Pictorially, one can think of this as ``shooting" from the IR and seeing where the flow stops in the UV. 

We start with the value of the parameter $g \simeq 0.84$. In this case no exotic RG flow is present. The flow starts in the IR at  $\phi_4$ and stops in the UV at  $\phi_3$. This is just an usual RG flow that interpolates smoothly between two adjacent AdS solutions. 

\begin{figure}[t!!!]
\centering
\includegraphics[width=.49\textwidth]{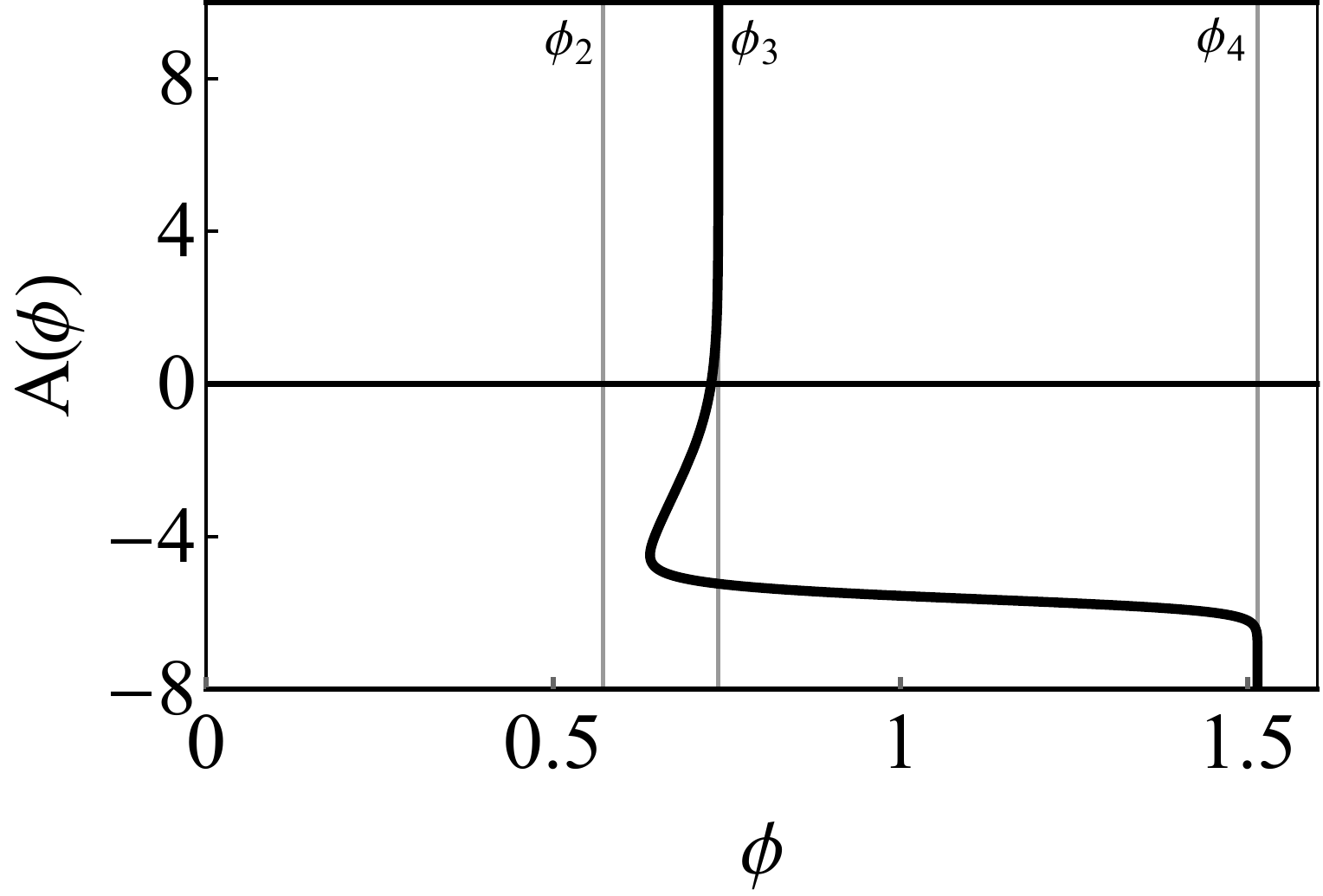}
\includegraphics[width=.49\textwidth]{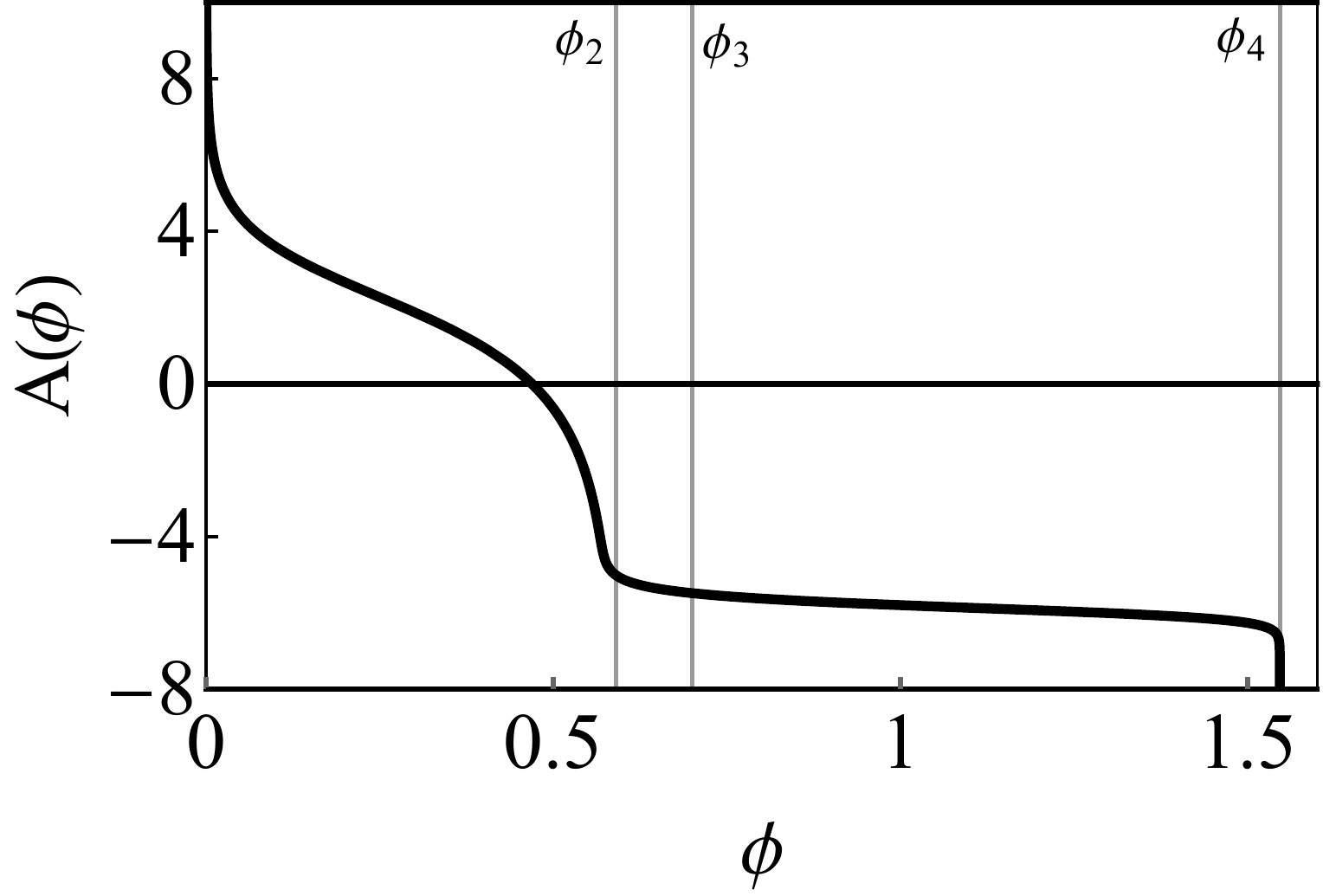}
\includegraphics[width=.49\textwidth]{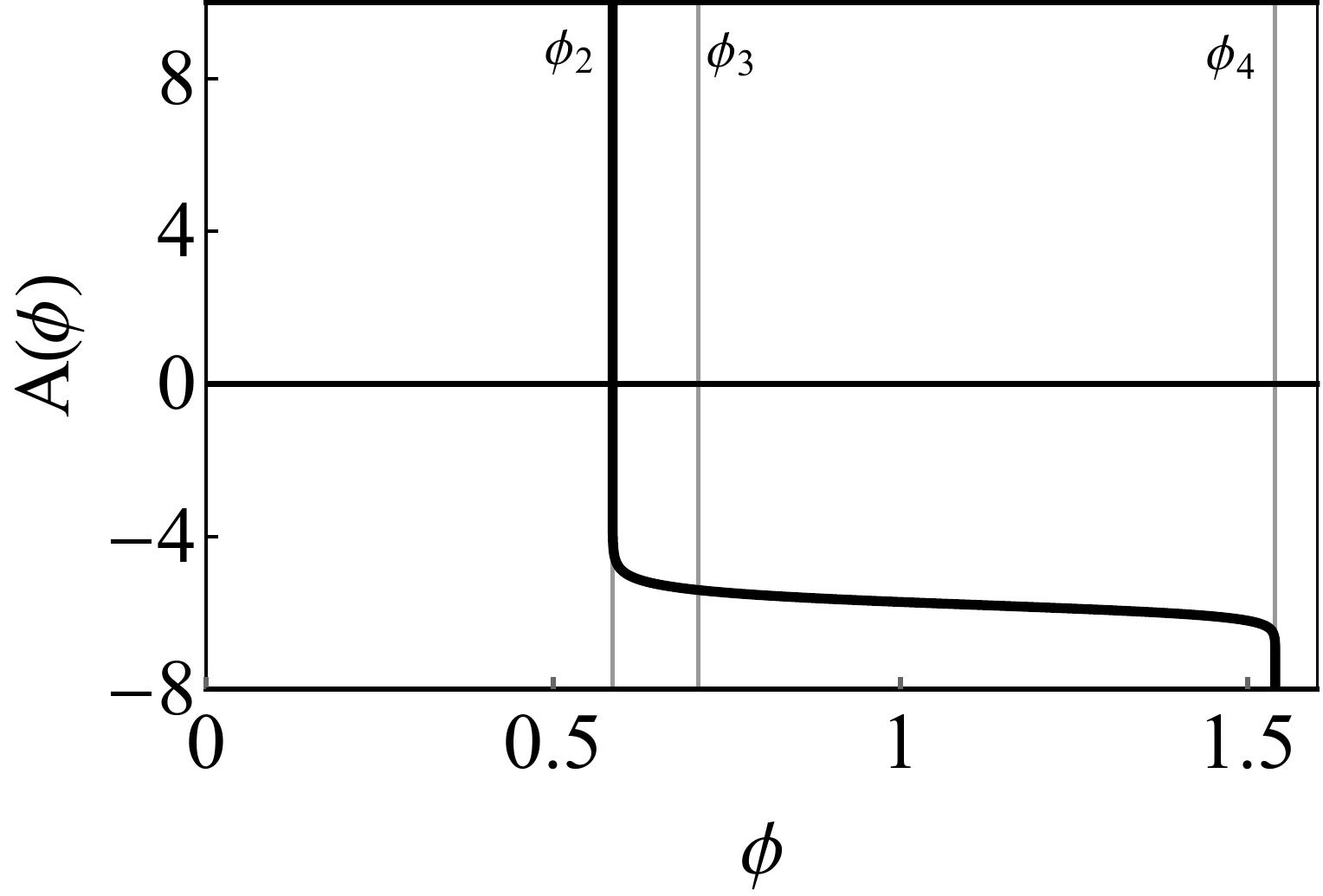}
\includegraphics[width=.49\textwidth]{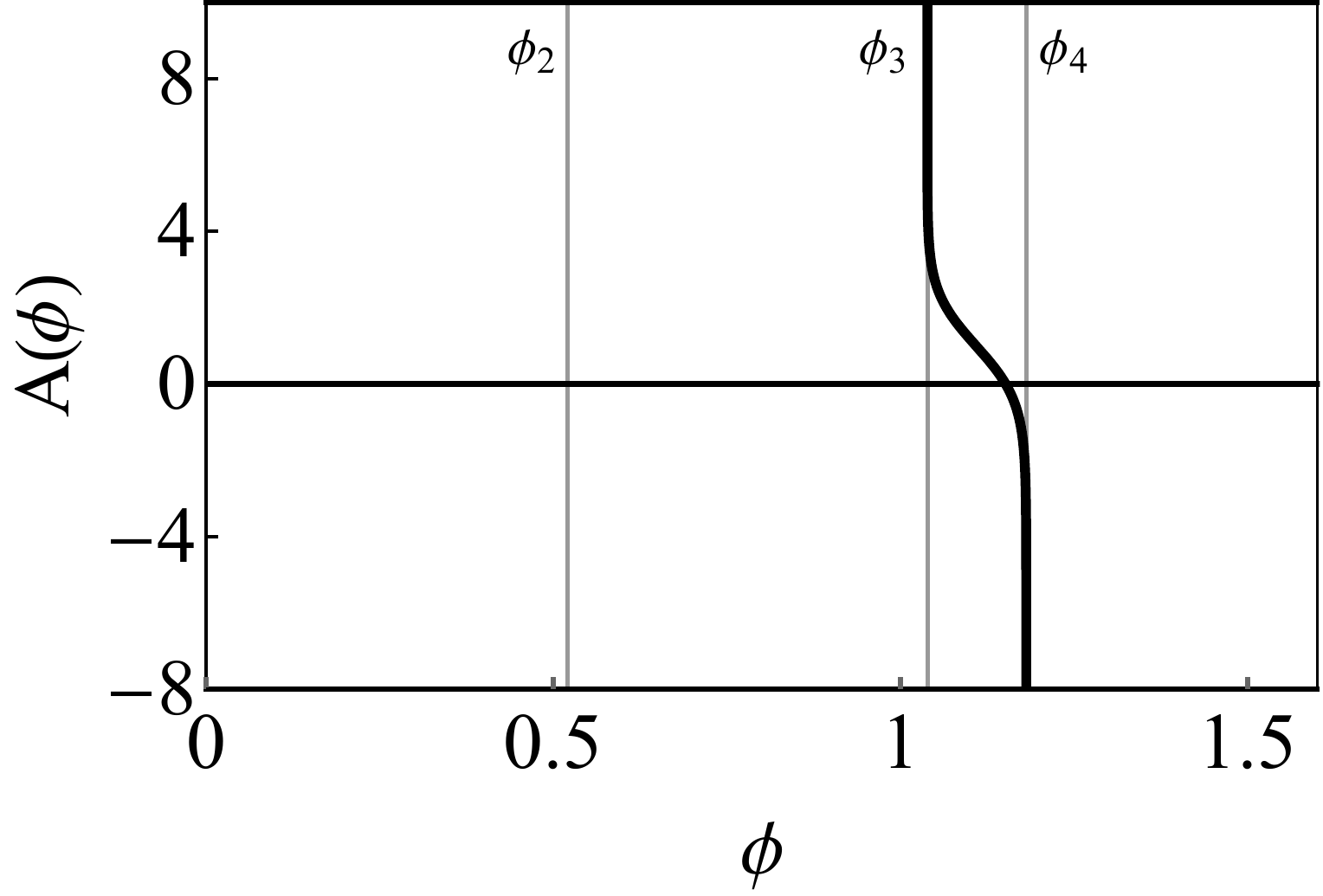}
\caption{\label{exoticRGflowsT0} \small
Continuous interpolation between different types of exotic RG flows. Top left:  Bouncing flow ($g=0.97$). As we increase the value of $g$, the bouncing point gets closer to $\phi_2$. Top right: Skipping flow ($g=0.99$). The bounce has become so large that it overpasses the minimum at $\phi_2$ and continues to $\phi_1=0$. Bottom left: Irrelevant VEV flow ($g=0.98589$). It corresponds to the limiting case between bouncing flow and skipping flow, or equivalently when the tip of the bounce matches $\phi_2$. Bottom right: Relevant VEV flow ($g=0.84143$). It corresponds to the limiting case between the bouncing flow and the usual, non-bouncing flow. Solutions are not necesarily normalized to source $\Lambda=1$ or VEV$=1$. Recall that for each value of $g$ the potential is different, so also the location of the extrema.}
\end{figure}

Next we consider $g\simeq0.97$. The solution starting at  $\phi_4$ bounces back before reaching  $\phi_3$, see \fig{exoticRGflowsT0}(top left). This exotic behavior is related to the global structure of the potential, and it is difficult to predict a priori based on a purely local analysis. It would be interesting to find some physical quantity that indicates when a given potential will have bouncing solutions before solving the equations. Nevertheless, after solving many different cases we have developed some intuition about when it does happen: the potential must be ``steep enough" between the maximum and the minimum, and also the minimum must be ``deep enough". 
From the viewpoint of the field theory, this bouncing flow corresponds to a \mbox{$\beta$-function} that has branch cuts \cite{Kiritsis:2016kog}.

If we now increase the value of $g$ in a continuous way up to $g\simeq0.99$, the minimum at $\phi_4$ gets deeper, the potential between $\phi_3$ and $\phi_4$ gets steeper, and the bounce gets larger. Eventually, the bounce is large enough to reach the minimum at $\phi_2$.  Then, instead of bouncing back, the flow continues and reaches $\phi_1$, see \fig{exoticRGflowsT0}(top right). Thus, we can think of the skipping flow as a bouncing flow in which the bounce is large enough to overpass the nearest minimum. From the viewpoint of the field theory, this flow corresponds to a 
$\beta$-function that overpasses two CFTs at $\phi_2$ and $\phi_3$, before reaching  $\phi_1$.

Now, if we fine-tune the parameter $g$ in such a way that we get the exact limiting case between the bouncing flow and the skipping flow, $g\simeq0.98589$, then the flow stops precisely at $\phi_2$. This corresponds to the irrelevant VEV flow,  see \fig{exoticRGflowsT0}(bottom left). At the minimum $\phi_2$ the dual scalar operator is irrelevant, and this flow correspond to turning on a VEV for that operator, while having vanishing source. VEV flows are non-generic in the sense that they require a fine-tuning of the potential. 

Finally, we can also fine tune the parameter $g$ to get the limiting case when the bounce starts forming, $g\simeq 0.84143$. This is precisely the relevant VEV flow, see \fig{exoticRGflowsT0}(bottom right). At the maximum $\phi_3$ the dual operator is relevant, and this flow corresponds to turning on a VEV while keeping the source equal to zero.

\begin{figure}[tbhp]
\centering
\includegraphics[width=.49\textwidth]{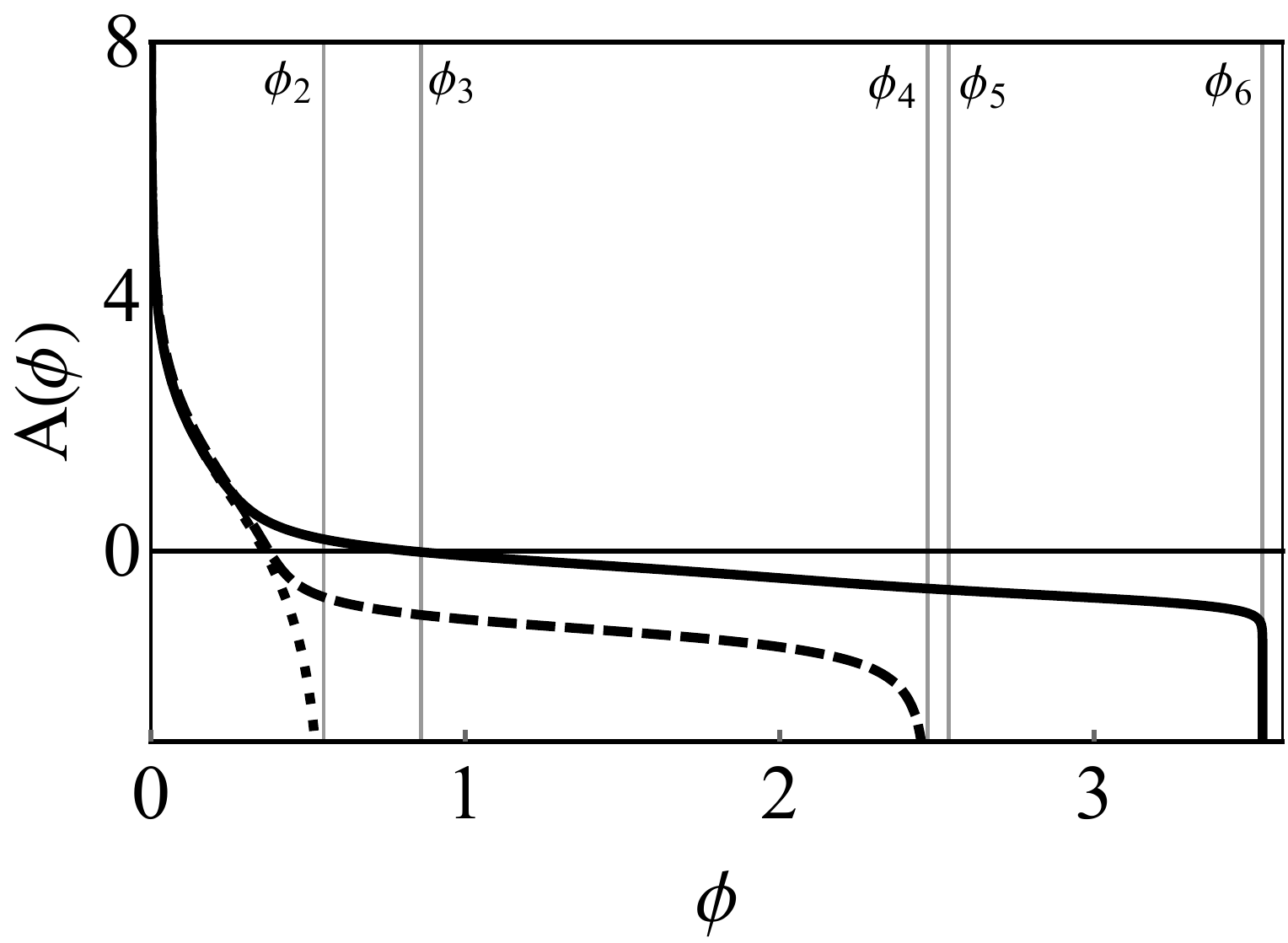}
\includegraphics[width=.49\textwidth]{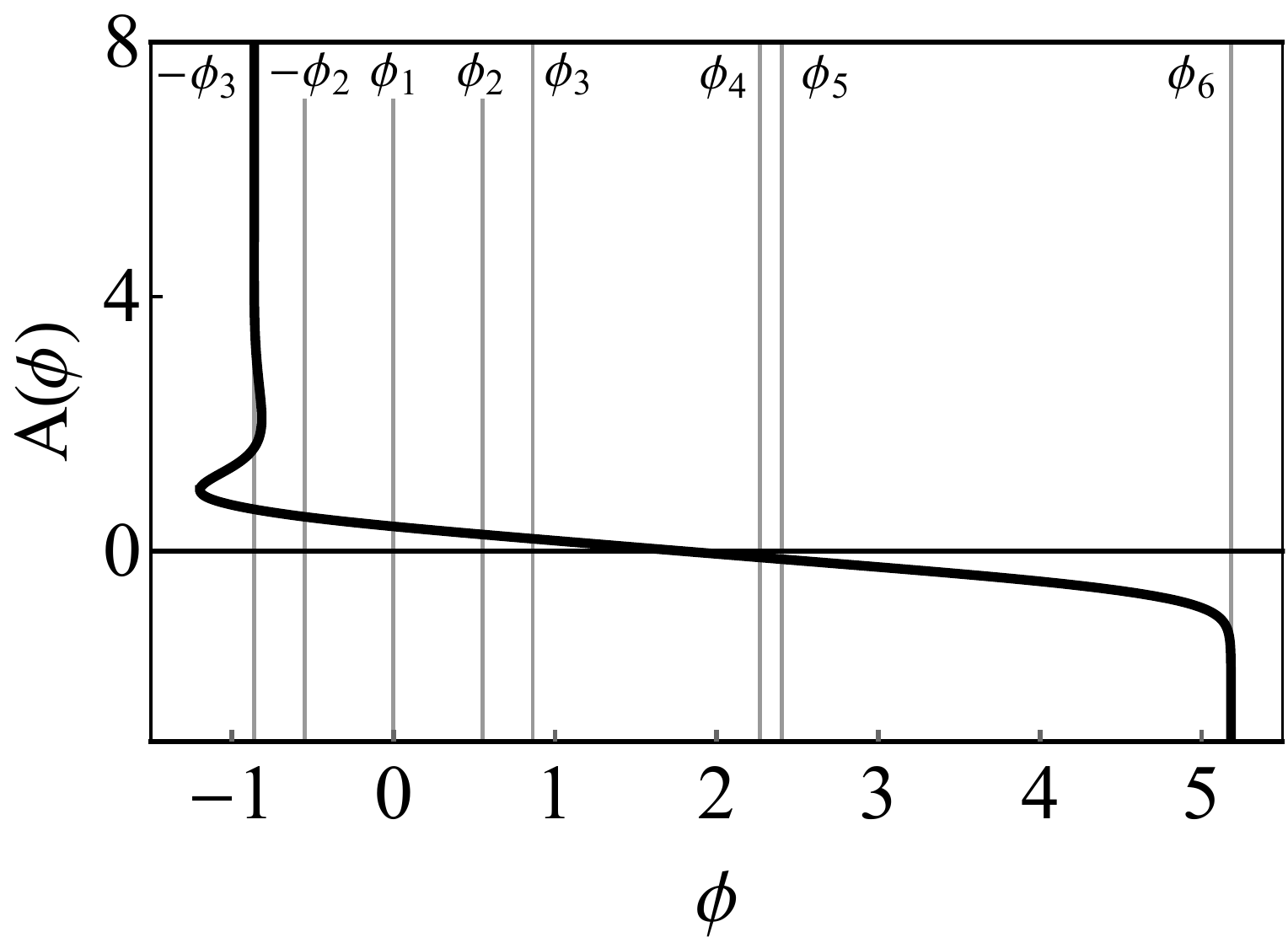}
\caption{\label{moreonexoticRGflowsT0}  \small
Left: The solid line is a skipping flow that skips four extrema. The dashed line is a flow that skips two extrema. Together with the flow ending at $\phi_2$, the dotted line, there are three different vacua for the CFT$_1$.  Right: an exotic RG flow that bounces twice and skips several extrema.}
\end{figure}

We close this section by noting that the different types of flows above can occur combined with one another for certain potentials. For illustration we will provide two examples. The first one is a skipping flow that skips not only two extrema but four. This arises as a solution with the following potential: 
\begin{equation}
L^2 V(\phi)=-3-\frac{3 \phi ^2}{8}+0.95 \phi ^4-0.8 \phi ^6+0.1903 \phi ^8-0.01785 \phi ^{10} +0.00057 \phi ^{12} \,.
\label{potential_skipping4}
\end{equation}
This potential has six extrema at $\phi_1 < \phi_2 < \phi_3 < \phi_4 < \phi_5 <\phi_6$, with $\phi_1$ a maximum, $\phi_2$ a minimum, etc.
We find solutions starting in the UV at $\phi_1=0$ and ending in the IR at $\phi_2$, $\phi_4$ and $\phi_6$, as shown in \fig{moreonexoticRGflowsT0}(left). Thus, in this case the CFT$_1$ has three non-degenerate vacua. Presumably this can  be extended to an arbitrary number of vacua by chosing the potential appropiately. 

The second example  is a flow that bounces twice and skips several extrema, see \fig{moreonexoticRGflowsT0}(right). We obtain this flow as a solution with  the following potential:
\begin{equation}
L^2 V(\phi)=-3-\frac{3 \phi ^2}{8}+0.95 \phi ^4-0.8 \phi ^6+0.1903 \phi ^8-0.0163 \phi ^{10} +0.00035 \phi ^{12} ~,
\label{potential_skipping_several}
\end{equation}
By fine tuning the parameters of this potential, one can also construct other interesting cases, for example, an irrelevant VEV flow that also skips several extrema. 

%%%%%%%%%%%%%%%%%%%%%%%%%%%%%%%%
%%%%%%%%%%%%%%%
%%%%%%%%%%%%%%%
%%%%%%%%%%%%%%%%%%%%%%%%%%%%%%%%

\section{Discussion}
We have studied exotic RG flows at non-zero temperature. In part, our results can be seen as a demonstration that the types of zero-temperature exotic RG flows listed in Section \ref{interpol} also exist at non-zero temperature. One relevant difference is the fact that the two cases that requiere fine tunning at $T=0$, the two VEV flows, do not require such fine tunning at non-zero temperature, since in this case the theory simply adjusts the ratio between the temperature and the VEV. 

The thermodynamics of the CFT$_1$ is sensitive to the entire form of the scalar potential on the gravity side between $\phi_1$ and $\phi_4$. Therefore it must encode much information about the other CFTs. The information about the CFT$_2$ and the CFT$_4$ is presumably encoded in the IR behaviour of the CFT$_1$ near its metastable and the stable vacua, respectively. The information about the CFT$_3$ may also be encoded in the CFT$_1$, albeit in a less obvious, possibly highly non-local way. It would be interesting to understand this in detail, perhaps along the lines of \cite{Gubser:2008ny}. A more intriguing possibility would be to be able to extend this analysis to the case in which the potential possesses a positive maximum at $\phi_3$. In this case the solution at this point becomes a dS spacetime instead of an AdS spacetime, but we have seen that nevertheless the thermodynamics of the CFT$_1$ is qualitatively the same. It would also be interesting to construct flows between this dS geometry and one of the AdS extrema, along the lines of \cite{Freivogel:2005qh}.

The gravity model that we have used is a bottom-up model. It would be interesting to find  top-down models that exhibit similar features. However, in  consistent truncations to five dimensions of ten- or eleven-dimensional supergravity that retain only one scalar it is uncommon that there is more than one extremum and, if so, the additional extrema are typically unstable because they violate the Breitenlohner-Freedman bound. 

It would also be interesting to extend our analysis to the case of multiple scalar fields, which at zero temperature has been analysed in \cite{Nitti:2017cbu}.

The CFT$_1$ in our model possesses two non-degenerate vacua. The one with higher energy is metastable. If one imagines that of our model may be the bosonic truncation of a supersymmetric model, then this provides 
a simple example of dynamical metastable supersymmetry breaking \cite{Intriligator:2006dd}. It would be interesting to use this model to study  time-dependent properties  such as the evolution of bubbles of the metastable vacuum.

%%%%%%%%%%%%%%%%%%%%%%%%%%%%%%%%
%%%%%%%%%%%%%%%
%%%%%%%%%%%%%%%
%%%%%%%%%%%%%%%%%%%%%%%%%%%%%%%%

\section*{Acknowledgements}
%%%%%%%%%%%%%%%%%%%%%%%%%%%%%%%%
%%%%%%%%%%%%%%%

We are extremely grateful to Jorge Casalderrey-Solana and Carlos Hoyos for extensive discussions. We also thank Roberto Emparan, Anton Faedo, Bartomeu Fiol, Elias Kiritsis and Francesco Nitti for discussions.  We are supported by grants SGR-2017-754, FPA2016-76005-C2-1-P, FPA2016-76005-C2-2-P, and Maria de Maeztu MDM-2014-0369 Unit of Research Excellence distinction. 

\appendix

\end{document}